%% file: pipi_ff.tex
\documentclass[11pt,a4paper]{article}
\pdfoutput=1   
\usepackage{amsmath,amsfonts,amssymb,amsbsy}
\usepackage{mathrsfs,latexsym}
\usepackage{graphicx}
\usepackage{xcolor}
\usepackage{booktabs}
\usepackage{multirow}
\usepackage{multicol}
\usepackage{subfig}
\usepackage{placeins}
\usepackage{blindtext}
\usepackage{bm}
\usepackage{tikz}
\usepackage{tikzscale}
\usetikzlibrary{patterns, calc, fit}
\usetikzlibrary{plotmarks, decorations.markings}

\usetikzlibrary{external}
\usepackage{alpha}
\usepackage{macros_alpha}
\renewcommand{\strange}{\mathrm{strange}}
\usebiblio{latticen}

\newcommand{\cm}{{\mathrm {cm}}}

\newcommand{\dvec}{\bm{d}}

\begin{document}

\input{title.tex}

\input{intro.tex}
\input{method.tex}

\input{results.tex}

\input{concl.tex}

\input{app.tex}

\vskip 0.3cm

\noindent
{\bf Acknowledgements}: 
We gratefully acknowledge communications with H. Meyer, 
J. R. Pelaez, C. Hanhart, M. T. Hansen, S. Schaefer, 
M. Dalla Brida, and F. Erben. 
 Special thanks to F. Knechtli for 
helping us secure computer resources used for this project and V. Koch for 
contributions at an early stage. 
BH was supported by Science Foundation Ireland under Grant No. 11/RFP/PHY3218. CJM acknowledges support from the U.S. NSF under award PHY-1613449 and through TeraGrid/XSEDE
resources provided by TACC, SDSC, and NICS under grant number TG-MCA07S017. 
Computing facilities
	were provided by the Danish e-Infrastructure Cooperation (DeIC) National HPC
Centre at the University of Southern Denmark. 
The authors wish to acknowledge the DJEI/DES/SFI/HEA Irish Centre for High-End Computing (ICHEC) for the provision of computational facilities and support.
We are grateful to our CLS 
	colleagues for sharing the gauge field configurations on
which this work is based.
We acknowledge PRACE for awarding us access to resource FERMI based in Italy
at CINECA, Bologna and to resource SuperMUC based in Germany at LRZ, Munich.
Furthermore, this work was supported by a grant from the Swiss National Supercomputing
Centre (CSCS) under project ID s384. We are grateful for the support received by the
computer centers.
The authors gratefully acknowledge the Gauss Centre for Supercomputing (GCS)
for providing computing time through the John von Neumann Institute for Computing
(NIC) on the GCS share of the supercomputer JUQUEEN at J\"{u}lich Supercomputing
Centre (JSC). GCS is the alliance of the three national supercomputing centres HLRS
(Universit\"{a}t Stuttgart), JSC (Forschungszentrum J\"{u}lich), and LRZ (Bayerische Akademie
der Wissenschaften), funded by the German Federal Ministry of Education and Research (BMBF) and the German State Ministries for Research of Baden-W\"{u}rttemberg (MWK),
Bayern (StMWFK) and Nordrhein-Westfalen (MIWF).
The USQCD QDP++ library~\cite{Edwards:2004sx} was used in designing 
the software used for the calculations reported here.



\end{document}

%% file: title.tex
\preprintno{%
CP3-Origins-2018-029 DNRF90\\
MITP/18-073\\
}

\title{%
	The $I=1$ pion-pion scattering amplitude and timelike pion 
	form factor from $\nf = 2+1$ lattice QCD 
}

\author[sdu]{Christian~Andersen}
\author[sdu]{John~Bulava}
\author[mainz]{Ben~H\"{o}rz}
\author[cmu]{Colin~Morningstar}

\address[sdu]{CP3-Origins, University of Southern Denmark, Campusvej 55, 5230 
Odense M, Denmark}
\address[mainz]{PRISMA Cluster of Excellence and Institut f\"{u}r Kernphysik,
University of Mainz, Johann-Joachim-Becher-Weg 45, 55099 Mainz, Germany}
\address[cmu]{Department~of~Physics, Carnegie~Mellon~University, 
Pittsburgh, PA~15213, USA}

\begin{abstract}
	The elastic $I=1$ $p$-wave $\pi\pi$ scattering amplitude is calculated
	together with the isovector timelike pion form factor using lattice 
	QCD with 
	$\nf=2+1$ dynamical quark flavors. Wilson clover ensembles generated by the 
	Coordinated 
	Lattice Simulations (CLS) initiative are employed at four lattice spacings 
	down to $a = 0.05\,\mathrm{fm}$, several pion masses down to 
	$m_{\pi} = 200\,\mathrm{MeV}$, and spatial volumes of extent 
	$L = 3.1-5.5\,\mathrm{fm}$. The set of measurements on these ensembles, 
	which is publicly available, 
	enables an investigation of 
	systematic errors due to the finite lattice spacing
	and spatial volume. 
	The $\pi\pi$ scattering amplitude is fit on each ensemble by a Breit-Wigner 
	resonance lineshape, while the form factor 
	is described better by a thrice-subtracted dispersion relation than the 
	Gounaris-Sakurai parametrization.
\end{abstract}


\maketitle

%% file: intro.tex
\section{Introduction}\label{s:intro}

Lattice QCD calculations of resonant two-hadron scattering amplitudes have 
improved markedly in recent years thanks to algorithmic 
advances~\cite{Peardon:2009gh,Morningstar:2011ka} and increased 
computing resources.\footnote{For recent reviews of the interplay
between lattice QCD calculations and current computer architectures, see 
Refs.~\cite{Boyle:2017vhi,Rago:2017pyb}.} Many calculations of the elastic 
$\pi\pi$ amplitude 
in the vicinity of the $\rho(770)$ exhibit sufficient statistical 
precision and energy resolution to determine the resonance 
parameters~\cite{Alexandrou:2017mpi,Bulava:2016mks,Fu:2016itp,Guo:2016zos,
Wilson:2015dqa,Feng:2014gba,Dudek:2012xn,Pelissier:2012pi,Aoki:2011yj,
Lang:2011mn,Feng:2010es},
while a few $K\pi$ calculations similarly map out the 
$K^{*}(892)$~\cite{Brett:2018jqw,Bali:2015gji,Wilson:2014cna,Prelovsek:2013ela}. 
 First coupled channel results have also appeared in 
 Refs.~\cite{Moir:2016srx,Briceno:2016mjc,Dudek:2016cru,Wilson:2014cna,Mohler:2013rwa}
 for the $a_0(980)$, $f_0(980)$, and $D^{*}_{s0}(2317)$ resonances. 
Resonant meson-meson amplitudes involving an external current have also been 
calculated in Refs.~\cite{Alexandrou:2018jbt,Briceno:2016kkp,Feng:2014gba}.
Compared to the meson-meson sector, calculations of resonant 
meson-baryon amplitudes are currently less advanced~\cite{Andersen:2017una,Lang:2016hnn,Lang:2012db}. A recent review 
of lattice calculations of scattering amplitudes can be found in 
Ref.~\cite{Briceno:2017max}.

The improvement in these calculations suggests that  
the  
quark-mass dependence of such amplitudes may be investigated 
quantitatively, providing valuable input to 
effective theories of low-lying hadron resonances as well as 
numbers at the physical point relevant for experiment. 
In order to obtain reliable results however, 
various systematic errors must be controlled. These include effects due to the 
finite lattice spacing and spatial volume inherent in lattice QCD simulations,
as well as systematics in the calculation of finite-volume two-hadron 
energies and matrix elements from which the amplitudes are determined. 

While lattice spacing effects are assessed in the usual way, the treatment of 
finite volume effects is
more subtle. Since 
real-time scattering amplitudes cannot be naively calculated from 
Euclidean-time lattice QCD simulations~\cite{Maiani:1990ca}, the method 
proposed by L\"{u}scher~\cite{Luscher:1990ux} is employed to infer two-to-two 
hadron scattering 
amplitudes from shifts of finite-volume two-hadron energies from their 
non-interacting values. This approach has been generalized to non-zero 
total momenta~\cite{Rummukainen:1995vs,Kim:2005gf}, non-zero 
spin~\cite{Gockeler:2012yj,Morningstar:2017spu,Briceno:2013lba,
Briceno:2014oea,Romero-Lopez:2018zyy,Woss:2018irj}, multiple 
coupled scattering channels~\cite{He:2005ey,Li:2012bi,Briceno:2014oea}, and amplitudes with 
an external current~\cite{Briceno:2015tza,Briceno:2015csa,Briceno:2012yi,
Hansen:2012tf,Meyer:2011um,Lellouch:2000pv}. Extending this approach above 
three-hadron thresholds has proven difficult and been applied to
Monte Carlo lattice data only in a toy scalar field theory~\cite{Romero-Lopez:2018rcb}.  It is, however, under active development~\cite{Mai:2018djl,Doring:2018xxx,Briceno:2018mlh,Hammer:2017kms,Hammer:2017uqm,
Briceno:2017tce,Hansen:2016ync,Hansen:2016fzj,
Hansen:2015zga,Hansen:2014eka, 
Polejaeva:2012ut,Briceno:2012rv}.
Recently-proposed alternatives to this finite-volume formalism for total decay 
rates can be 
found in Refs.~\cite{Hansen:2017mnd,Hashimoto:2017wqo}.

The relation discussed above between finite-volume energy shifts and 
real-time two-to-two scattering amplitudes can be written 
as~\cite{Morningstar:2017spu}
\begin{align}\label{e:det} 
	\mathrm{det} [\tilde{K}^{-1}(E_{\cm}) - B^{(\Lambda, \boldsymbol{d})}(E_{\cm})] = 0
\end{align}
where $E_{\cm}$ is the finite-volume two-hadron energy in the center-of-mass 
frame, $\tilde{K}$ is proportional to the infinite-volume $K$-matrix, and $B$ is a known matrix 
encoding the effect of the finite volume. This determinant is 
block diagonalized so that Eq.~\ref{e:det} describes finite-volume energies 
in a single irreducible representation (irrep) $\Lambda$ of the little group 
for a particular class of total momenta $\boldsymbol{d} = (L/2\pi) \boldsymbol{P}_{\mathrm{tot}}$, where $\boldsymbol{d}$ is a vector of integers.  
The determinant is taken over total 
angular momentum ($J$), total spin ($S$), all coupled two-hadron scattering channels, and an 
index enumerating the possibly multiple occurrences of a partial wave in irrep
$\Lambda$.

The determinant condition in Eq.~\ref{e:det} holds up to corrections which 
are exponentially suppressed in the spatial extent $L$. However, unlike 
finite volume corrections to single-hadron observables, the fall-off of these 
residual exponential finite volume effects may in principle be different
than $m_{\pi}$. The `rule of thumb' $m_{\pi}L \gtrsim 4$ which is 
usually applied in single-hadron calculations to ensure that finite-volume 
effects are at the percent level must be re-investigated in the 
context of scattering amplitudes. There exists therefore a hierarchy 
of finite volume effects: those described by Eq.~\ref{e:det} are polynomial  
in $L^{-1}$ (and constitute the `signal') while Eq.~\ref{e:det} holds 
only up to (unwanted) 
terms exponential in $L$. 

As a benchmark amplitude suitable for an investigation of these systematic 
effects, we consider here
the elastic $I=1$ $p$-wave
pion-pion scattering amplitude relevant for the $\rho(770)$.   
In order to extrapolate to the physical point and continuum, a range of pion 
masses $m_{\pi}=200-280\,\mathrm{MeV}$ and lattice spacings 
$a = 0.050-0.086\,\mathrm{fm}$ are employed. Such extrapolations have been 
performed recently using chiral perturbation 
theory and its extensions~\cite{Djukanovic:2004mm,Djukanovic:2009zn,
Nebreda:2010wv,
Djukanovic:2014rua,
Bavontaweepanya:2018yds,Hu:2017wli,Bolton:2015psa,
 Liu:2016wxq,Liu:2016uzk,MartinezTorres:2017bdo,Guo:2016zep,Guo:2018kno}
 and treat all scattering data simultaneously. As these extrapolations are 
 somewhat involved and have not yet treated cutoff effects, we 
present the determination of the amplitudes only in this work and leave 
extrapolation to 
the physical quark masses and continuum for the future. Nonetheless, in our 
results the residual finite volume and cutoff effects are evidently small 
compared to the statistical errors. 

We calculate scattering amplitudes over an energy range 
$E_{\cm} \in [ 2m_{\pi}, E_{\rm max}]$. 
Since Eq.~\ref{e:det} applies below $n>2$ hadron thresholds, $E_{\rm max}$ 
is reduced as $m_{\pi}$ is lowered. Because 
of the chiral trajectory employed in this work (which is discussed in 
Sec.~\ref{s:cls}), for the heaviest quark masses the lowest inelastic 
threshold is $\bar{K}K$, while the lowest $n>2$ hadron threshold is $4\pi$. Although levels in the range $2m_{\rm K} < E_{\cm} < 4m_{\pi}$ could be treated using 
Eq.~\ref{e:det} with a coupled-channel $K$-matrix, we nonetheless impose a 
restriction to elastic scattering, namely $E_{\rm max} = \min(4m_{\pi}, 2m_{\rm K})$.  

In addition to this $\pi\pi$ scattering amplitude, we also calculate the 
$I=1$ timelike pion form factor, which encodes the coupling of an external 
(timelike) photon to two pions in an isovector configuration.  
Phenomenologically, it can be extracted from 
$\mathrm{e}^+\mathrm{e}^- \rightarrow \mathrm{hadrons}$ and hadronic 
$\tau$-decays~\cite{Jegerlehner:2009ry} and is of particular relevance for 
the hadronic vacuum polarization (HVP), a leading source of theoretical 
uncertainty in the anomalous magnetic moment of the muon 
$(g-2)_{\mu}$~\cite{Meyer:2018til,Giusti:2018mdh}. 
Using the optical theorem, the imaginary part of the HVP can be related to 
\begin{align}
	R_{\mathrm{had}}(s) = \sigma(\mathrm{e}^{+}\mathrm{e}^{-} 
	\rightarrow \mathrm{hadrons} )/\frac{4\pi\alpha_{\mathrm{em}}(s)^2}{3s},
\end{align}
where $\sigma(\mathrm{e}^{+}\mathrm{e}^{-} 
	\rightarrow \mathrm{hadrons} )$ is the total cross section, 
	$\alpha_{\mathrm{em}}$ the electromagnetic coupling, and 
	$s=E_{\mathrm{cm}}^2$ the usual 
	Mandelstam variable. In the elastic region  
	$R_{\mathrm{had}}$ is given by the two-pion contribution
	\begin{align}
		R_{\mathrm{had}}(s) = \frac{1}{4}\left( 1 - \frac{4m_{\pi}^2}{s}\right)^{\frac{3}{2}}|F_{\pi}(s)|^2,
	\end{align}
which contains the timelike pion form factor $F_{\pi}(s)$. 
The phase of this form factor is fixed by Watson's theorem, so we are interested
in the amplitude only here. Furthermore, we work in the isospin limit, where
electromagnetic interactions are ignored and $m_{\rm u}=m_{\rm d}$. Because of this, 
the elastic region persists up to either $s=4m_{\rm K}^2$ or $s=16m_{\pi}^2$. 

Although a precise determination of this form factor is phenomenologically 
desirable, there exists only the pioneering determination of 
Ref.~\cite{Feng:2014gba} which employs a single lattice spacing, heavier 
quark masses, and a (single) smaller physical volume than this work.
It is therefore imperative to also investigate lattice spacing 
and finite volume effects for this quantity, the former of which may be 
affected by renormalization and $\mathrm{O}(a)$-improvement of the 
electromagnetic current. 

In addition to its phenomenological impact, the timelike pion form factor 
is an important stepping stone toward more complicated resonance 
photoproduction amplitudes. Such amplitudes are relevant for ongoing and 
future experiments which photoproduce resonances. An additional step in this 
direction is the  $\pi \gamma \rightarrow \pi\pi$ amplitude studied using 
lattice QCD in Refs.~\cite{Alexandrou:2018jbt,Briceno:2016kkp}. However, the 
timelike pion 
form factor calculated here does not require disconnected flavor-singlet 
Wick contractions, which are ignored in Refs.~\cite{Alexandrou:2018jbt,
Briceno:2016kkp}.

Preliminary work toward the results reported here is found in Ref.~\cite{Bulava:2015qjz}.
The remainder of this paper is organized as follows. For completeness we 
review the  
 gauge field ensembles, methods for calculating finite-volume two-pion energies and matrix elements, and their relation to 
infinite-volume scattering amplitudes in Sec.~\ref{s:meth}. 
Results are given in Sec.~\ref{s:res} and
 conclusions in Sec.~\ref{s:concl}. 

%% file: method.tex
\section{Lattice QCD Methods}\label{s:meth} 

The subset of CLS ensembles used in this work is discussed in 
Sec.~\ref{s:cls} and application of the stochastic LapH method for all-to-all 
quark propagation in  
Sec.~\ref{s:slaph}. The analysis strategy used to extract the required 
finite volume energies and matrix elements from temporal correlation functions 
is contained in Sec.~\ref{s:ana}, while the relation between 
finite-volume quantities and infinite-volume scattering amplitudes is given 
in Sec.~\ref{s:form}.

\subsection{Gauge field ensembles}\label{s:cls} 

The ensembles of gauge field configurations employed here are 
from the Coordinated Lattice Simulations (CLS) initiative 
and are presented in 
Refs.~\cite{Bruno:2014jqa,Bali:2016umi}. They employ 
the  tree-level improved L\"{u}scher-Weisz gauge action~\cite{Luscher:1984xn}
and non-perturbatively $\mathrm{O}(a)$-improved Wilson 
fermions~\cite{Bulava:2013cta}. 
Open boundary conditions~\cite{Luscher:2011kk} are implemented in the temporal
direction. 
Although these boundary conditions were adopted to reduce 
autocorrelation times of the global topological charge, they also influence finite-temporal-extent  effects in temporal 
correlation functions. 

Contributions to two-hadron correlation functions where the hadrons 
 propagate in opposite temporal directions, which for identical 
 particles and zero total momentum yield a constant in time~\cite{Detmold:2008yn,Prelovsek:2008rf,Feng:2009ij,Dudek:2012gj}, are present 
 with periodic boundary conditions but absent in this setup.
 Therefore, for large temporal extent $T$ and if both interpolators are far from the boundaries, all two-point correlation functions with open temporal boundary 
 conditions have the form  
\begin{align}\label{e:bnd}
	\lim_{T\rightarrow \infty \atop 
	t_0, (T-t_{\rm f}) \rightarrow \infty } C_{T}(t_0,t_{\rm f}) = C(t_{\rm f}-t_0) \times 
	 \left\{ 1 + 
	 \mathrm{O}(\mathrm{e}^{-E_0 t_{\mathrm{bnd}}})\right\},
 \end{align}
 where $C_{T}(t_0,t_{\rm f}) = \langle \mathcal{O}(t_{\rm f})\bar{\mathcal{O}}(t_0)\rangle_T$ is the correlator with open boundaries of extent $T$, 
 $C(t) = \langle \mathcal{O}(t) \bar{\mathcal{O}}(0) \rangle$  
 the correlator in the $T \rightarrow \infty$ limit,  $E_0$ the lightest state 
 with vacuum quantum numbers and 
 $t_{\mathrm{bnd}} = \min(t_0,T-t_{\rm f})$ the minimal distance from an interpolator 
 to the temporal boundaries. Since presumably $E_0 \approx 2m_{\pi}$, if 
 $m_{\pi} t_{\mathrm{bnd}} \gtrsim 2$ then the exponential corrections
 in Eq.~\ref{e:bnd} are parametrically similar to exponentially suppressed 
 finite-volume effects in single-hadron energies.

\begin{figure}
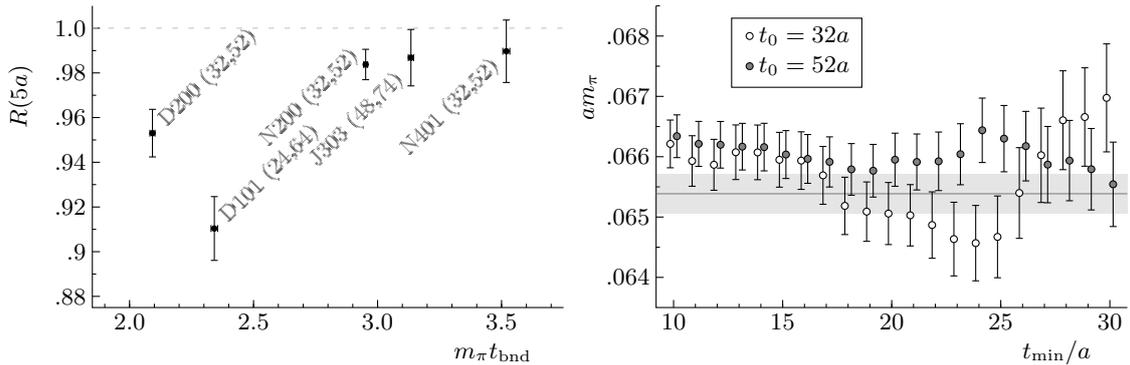

	 \includegraphics{tikz-figures/tbndRatio.tikz}
	 \includegraphics{tikz-figures/d200_srcTimeCheck.tikz}
	 \caption{\label{f:tmin_pion} \emph{Left}: $R(5a)$ from Eq.~\ref{e:trat} 
	 for all $(\frac{t_0}{a},\frac{t_0'}{a})$ pairs on ensembles where multiple source times 
	 are employed. \emph{Right}: $\tmin$-plot from single-exponential fits (which
	 ignore boundary effects) to the zero-momentum single-pion correlator
	 over the range $[\tmin,\tmax]$ for each $t_0$ individually on the D200 ensemble.  
	 The pion mass extracted from the $t_0$-averaged correlator is shown by the error band.
	 } 
 \end{figure}

 While correlation functions are 
 affected by temporal boundary conditions, the  
 transfer matrix (and therefore also the spectrum) is unaffected. 
 Although having an interpolating operator near a temporal boundary 
 does not change its quantum numbers, we are after excited 
 states and employ generalized eigenvalue methods requiring
 hermitian correlation matrices. The source and sink interpolators therefore 
 must not be significantly affected by the temporal boundaries in order to 
 maintain hermiticity. 
 To this end, we always choose
 a minimum distance to the boundary ($t_{\mathrm{bnd}}$) of at least 
 $m_{\pi}t_{\mathrm{bnd}} \gtrsim 2$.  
 As in Ref.~\cite{Bulava:2016mks}, our resulting insensitivity to finite-$T$ 
 effects can be demonstrated using the single-pion correlation function. 
Fitting this correlation function to a single exponential ignores 
contributions from the temporal boundaries. Fits of this type on a 
single ensemble are shown in Fig.~\ref{f:tmin_pion}, where the fitted energy 
is shown to be insensitive to the source interpolator position $t_{0}$.
Insensitivity to $t_0$ in our most precisely determined correlation function 
suggests that temporal boundary effects may be neglected in subsequent fits.  

 Another measure of finite-$T$ effects is the ratio 
 \begin{align}\label{e:trat}
	 R_{t_0,t_0'}(t) = \frac{C_T(t_0,t+t_0)}{C_T(t_0',t+t_0')}
 \end{align}
which under the asymptotic assumptions of Eq.~\ref{e:bnd} receives corrections 
to unity of $\mathrm{O}(\mathrm{e}^{-E_0t_{\rm bnd}})$. $R(5a)$ is also 
shown in Fig.~\ref{f:tmin_pion} for various $(t_0,t_0')$ pairs on 
all ensembles with multiple source times. This ratio shows significant deviations from unity for 
$m_\pi t_{\rm bnd} \lesssim 3$, despite no observable difference in the fitted 
energies. While the deviation of $R_{t_0,t_0'}(t)$ from unity 
in the single-pion correlator suggests that averaging over source times may 
affect the hermiticity of correlation matrices, such deviations are not 
visible in two-pion correlation functions. For the D200 ensemble the ratio with $(t_0/a,t_0'/a) = (32,52)$, shown for the pion in Fig.~\ref{f:tmin_pion} left 
panel as the left-most point, is consistent with unity for the single-$\rho$ meson correlator at rest and the two-pion correlator (each with a single unit of momentum) in the same channel.     

 While we omit a complete discussion of algorithmic details used in 
 configuration 
 generation, some aspects are relevant for the analysis of correlation 
 functions measured on these ensembles. 
 The CLS ensembles employ twisted-mass 
 reweighting~\cite{Luscher:2008tw} for the 
 degenerate light quark doublet and use the RHMC 
 algorithm~\cite{Clark:2006fx} for the 
 strange quark determinant. One reweighting factor ($W_0$) is  
 used to change the light 
 quark action to clover Wilson, while another ($W_1$) 
 corrects for the RHMC approximation. 
 Efficient evaluation of these reweighting factors is discussed in 
 Ref.~\cite{Bruno:2014jqa}. 
Measurements of primary observables must be multiplied by the corresponding 
re-weighting factors on each configuration according to 
\begin{align}\label{e:rw}
	\langle A \rangle = \frac{\langle A W\rangle_W}{\langle W\rangle_W} 
\end{align}
where $W=W_0W_1$ and $\langle \dots \rangle_W$ denotes an ensemble average 
with respect to the simulated action. The denominator of Eq.~\ref{e:rw} must
also be taken into account in any resampling procedure used to estimate 
statistical errors or covariances. If both the twisted mass parameter and the range and degree of the rational approximation are chosen appropriately, these 
reweighting factors are typically close to unity. However, we observe 
anomalously large fluctuations in the reweighted zero-momentum 
single-pion correlator for a single 
source time on each of the C101 and D101 ensembles. These ensembles have the 
lightest 
quark mass at the coarsest lattice spacing and large fluctuations may
indicate an inefficient choice of the simulated action. Data 
on these two source times are removed from the final analysis and are not included in Tab.~\ref{t:ens}.  

There are several possibilities 
for the trajectory of $m_{\rm s}$ as $m_{\rm l} = m_{\rm u} = m_{\rm d}$ is lowered toward its physical value. Quark masses on the CLS ensembles employed here are tuned to lie on a chiral 
trajectory with $\mathrm{tr}\, M_{\rm q} = \mathrm{const}.$, where $M_{\rm q}$ is the bare quark mass matrix $M_{\rm q} = \mathrm{diag}(m_{\rm u},m_{\rm d},
m_{\rm s})$, in order to reduce the quark 
mass dependence of certain renormalized quantities. 
An additional chiral trajectory in which $m_{\rm s} = \mathrm{const}.$ is 
presented in Ref.~\cite{Bali:2016umi}. While it is interesting to investigate 
the 
quark-mass dependence of scattering amplitudes on 
both chiral trajectories, we present results on the $\mathrm{tr} \, M_{\rm q} = 
\mathrm{const.}$ trajectory only.

As discussed in Ref.~\cite{Bruno:2016plf}, fixing the trace of the bare mass 
matrix is not equivalent to
fixing the sum of renormalized quark masses. There a Taylor expansion is 
employed to slightly shift the quark masses in order to satisfy $\phi_4 = 8t_0 (m_{\rm K}^2 + \frac{1}{2}\mpi^2)=\mathrm{const.}$, which is not performed here.  
At the coarsest lattice spacing, imposing $\mathrm{tr}\, M_{\rm q} =\mathrm{const}.$ 
results in deviations of less than $5\%$ of 
\,$\mathrm{tr}\, M^{\rm R}/(\mathrm{tr}\, M^{\rm R})^{\rm symm}$ (where the trace in the denominator is evaluated at the symmetric point $m_{\rm u}=m_{\rm d}=m_{\rm s}$) from unity at the
lightest pion masses 
considered in Ref.~\cite{Bruno:2016plf}. This small deviation from our desired chiral trajectory presumably has little effect on the observables 
considered here. 

Properties of the CLS ensembles used in this work are given in Tab.~\ref{t:ens}, which also contains $\tau_{\mathrm{meas}}$, the separation in molecular 
dynamics units (MDU) between our measurements of hadronic correlation functions, and $t_{\mathrm{bnd}}$, the 
minimum distance from an interpolator to a temporal boundary. 
The timelike pion form factor is not determined on the coarsest lattice spacing.
\begin{table}
	\centering
	\begin{tabular}{cccccccc}
		\toprule
		ID & $\beta$ & $a \,\mathrm{[fm]}$ & $L^3 \times T$  & $\mpi,\, m_{K} \,\mathrm{[MeV]}$ & $\tau_{\mathrm{meas}} \, \mathrm{[MDU]}$ &
		$N_{\mathrm{conf}}$ &
		$m_{\pi}t_{\mathrm{bnd}}$   \\ 
		\midrule
		C101 & $3.4$ & $0.086$   & $48^3\times96\hphantom{8}$ & $220,\,470$ & $8$  & $300$ & 
		$2.5$   \\
		D101 & & & $64^3\times128$ &  & $8$ & $303$ &  $2.3$  \\
		\midrule
		N401 & $3.46$ & $0.076$ & $48^3\times128$ & $280,\, 460$ & $16$ & $274$ & $3.5$  \\
		\midrule
		N200 & $3.55$ & $0.064$ & $48^3\times128$ & $280,\,460$ & $8$  & $854$ & $3.0$   \\
		D200 & &  & $64^3\times128$ & $200,\, 480$  & $8$ & $558$  & $2.1$   \\
		\midrule
		J303 & $3.7$ & $0.050$ & $64^3\times192$ & $260,\,470$ & $16$ & $328$ & $3.1$  \\
	\end{tabular}
	\caption{\label{t:ens} Parameters of the CLS ensembles used in this work.
	The timelike pion form factor is not determined on the coarsest lattice 
	spacing.
	After the ensemble ID in the first column, we list 
	the gauge coupling, lattice spacing and dimensions, pseudoscalar meson 
	masses, the separation between correlation function measurements in molecular dynamics units (MDU), the number of such measurements, and the minimum 
	distance from an interpolator to a temporal boundary.
	} 
\end{table}
A more precise scale determination can be found in Tab.~3 of 
Ref.~\cite{Bruno:2016plf}, while pseudoscalar meson masses and decay constants 
can be found in Tab.~2 of that work.

As discussed above, open temporal boundary conditions are employed to 
decrease the integrated autocorrelation time of the global topological charge.
However, there is still a significant amount of 
autocorrelation present in some observables on the CLS ensembles. A method to estimate 
statistical errors in the presence of 
large autocorrelations is outlined in 
Refs.~\cite{Schaefer:2010hu,Wolff:2003sm}. 
This involves propagating the errors linearly, a method which may not be 
suitable for our purposes given the non-linear nature of the $B$-matrix 
elements given in Eq.~\ref{e:det} and discussed further
in Sec.~\ref{s:ana}. Therefore, we simply `bin' our correlator measurements and 
employ the bootstrap procedure with $N_{B}=800$ bootstrap samples. Although
no statistically 
significant autocorrelations are observed in any of our correlation functions, 
the largest integrated autocorrelation times ($\tau_{\rm int}$) measured on 
these ensembles span the range $\tau_{\rm int} \approx 30-150$ for 
$\beta = 3.4-3.7$, respectively~\cite{Bruno:2014jqa}.

\subsection{Correlation function construction}\label{s:slaph} 

Since we employ two-pion interpolators in which each pion is projected to definite momentum,  quark propagators between all space-time points are required. We employ 
the stochastic LapH method to estimate such all-to-all propagators and efficiently 
construct correlation functions~\cite{Morningstar:2011ka}. Based on 
Ref.~\cite{Peardon:2009gh}, this method endeavors to make 
all-to-all propagators tractable by considering quark propagation between 
a low-dimensional subspace defined by the lowest $N_{\rm ev}$ modes of the 
three-dimensional gauge-covariant Laplace operator, hereafter referred to as the 
`LapH subspace'. This projection is a form of quark 
smearing, with an approximately Gaussian spatial profile and width 
controlled by the $N_{\rm ev}$-th eigenvalue. In order to maintain a constant width, $N_{\rm ev}$ must be scaled  
proportionally to the spatial volume. 

This smearing procedure enables more efficient stochastic estimation schemes
by employing noisy combinations of Laplacian eigenvectors. 
It was determined in Ref.~\cite{Morningstar:2011ka} that (at least for the 
range of spatial 
volumes considered there) with a moderate level of dilution the quality of the 
stochastic estimator 
remains constant as the volume is increased while maintaining a fixed 
number of dilution projectors. This work further demonstrates that the quality 
of the stochastic LapH estimator does not degrade for even larger volumes. 
Without significantly 
increasing the number of dilution projectors, we obtain precise results for 
scattering amplitudes with stochastic LapH on spatial volumes up to 
$V=(5.5\,\mathrm{fm})^3$.

\begin{table}
	\centering 
	\begin{tabular}{cccccccc}
\toprule
		ID & $(\rho, n_{\rho})$  &$N_{\mathrm{ev}}$ & dilution & 
		$N^{\mathrm{fix}}_{\rm R}$ & $N^{\mathrm{rel}}_{\rm R}$ & 
		$N_{t_0}$ & $N_{\rm D}$   \\
\midrule
		C101 & $(0.1,20)$ & $392$ &$(\mathrm{TF,SF,LI16})_{\rm F}\,(\mathrm{TI8, SF, LI16})_{\rm R}$ & $6$ & $2$ & $1$ & $1408$  \\ 
		D101 &  & 928 &$(\mathrm{TF,SF,LI16})_{\rm F}\,(\mathrm{TI8, SF, LI16})_{\rm R}$ & $6$ & $2$ & $2$ & $1792$ \\ 
		\midrule
		N401 & $(0.1,25)$ & $320$ &$(\mathrm{TF,SF,LI16})_{\rm F}\,(\mathrm{TI8, SF, LI16})_{\rm R}$ & $5$ & $2$ & $2$ & $1664$  \\ 
		\midrule 
		N200 & $(0.1,36)$ & $192$ & $(\mathrm{TF,SF,LI8})_{\rm F}\,(\mathrm{TI8, SF, LI8})_{\rm R}$ & $5$ & $2$ & $2$ & $832$ \\ 
		D200 &   & $448$  &$(\mathrm{TF,SF,LI8})_{\rm F}\,(\mathrm{TI8, SF, LI8})_{\rm R}$ & $5$ & $2$ & $2$ & $832$   \\ 
		\midrule
		J303 & $(0.1,60)$ & $208$ &$(\mathrm{TF,SF,LI8})_{\rm F}\,(\mathrm{TI16, SF, LI8})_{\rm R}$ & $5$ & $2$ & $3$ & $1504$ \\
		\bottomrule
	\end{tabular}
	\caption{\label{t:laph}Parameters of the stochastic LapH implementation used 
	in this work.  $(\rho,n_{\rho})$ are the stout link smearing 
	parameters, $N_{\mathrm{ev}}$ the number of Laplacian eigenvectors, $N_{\rm R}$ the number of 
	independent noise sources, $N_{t_0}$ the number of source times for fixed 
	quark lines, and $N_{\rm D}$ the total number of light quark Dirac matrix 
	inversions per gauge configuration. Notation for the dilution scheme is explained in the text.}
\end{table}
In the stochastic LapH framework,  $N_{\rm R}$ stochastic sources  $\{\rho_r\}$  
are introduced in 
 time, spin, and Laplacian eigenvector indices. These sources are diluted by 
 specifying  
$N_{\mathrm{dil}}$ complete orthogonal dilution projectors $\{P_b\}$ so that 
an 
unbiased estimator of the smeared-smeared all-to-all quark propagator is 
furnished by 
\begin{align}\label{e:laph}
	\mathcal{Q} (y,x) \approx \frac{1}{N_{\rm R}} \sum_{r=1}^{N_{\rm R}} \sum_{b = 1}^{N_{\mathrm{dil}}} \varphi_{rb} (y) \, \varrho^{\dagger}_{rb}(x), 
\end{align}
where  
$\varrho_{rb} =  V_{\rm s} P_b\, \rho_r$ is the smeared stochastic source, 
$\varphi_{rb} = \mathcal{S} \, Q \, \varrho_{rb}$ the smeared sink, 
$Q$ the quark propagator, 
and $\mathcal{S}=V_{\rm s} V_{\rm s}^{\dagger}$ the smearing operator which projects onto the LapH subspace. 
To 
date, only schemes where dilution in each of these indices is done  
independently have been employed. 
A common strategy is to 
interlace $n$ dilution projectors uniformly (denoted 
$\mathrm{I}n$) in the index in question. The `full' dilution limit 
(denoted `F') is recovered if $n$ is equal to the 
total dimension of the index. Full specification of a dilution scheme 
therefore specifies a prescription in each of time, spin, and Laplacian 
eigenvector space. For example $(\mathrm{TF},\mathrm{SF},\mathrm{LI8})$ refers to full dilution in time 
and spin, and eight dilution projectors interlaced uniformly among the Laplacian 
eigenvectors.

As discussed in
Ref.~\cite{Morningstar:2011ka}, it is typically beneficial to employ 
different dilution schemes for `fixed' quark propagators (denoted by 
the subscript `F'), where $x_0 \ne y_0$  
 and `relative' quark propagators (denoted `R') where $x_0=y_0$. 
We therefore employ either full or interlace dilution in time, full dilution in
spin, and interlace dilution in eigenvector space.  
The dilution scheme and other parameters of the stochastic LapH algorithm 
employed here are given in 
Tab.~\ref{t:laph}. 

Tab.~\ref{t:laph} also contains information on the LapH subspace and thus 
the smearing operator $\mathcal{S}$ applied to quark fields in our 
interpolating 
operators. Before calculating eigenvectors, the gauge link field entering the 
covariant 3-D Laplace operator is stout smeared~\cite{Morningstar:2003gk}. 
The stout smearing parameters $(\rho,n_{\rho})$ together with the number 
of retained eigenvectors $N_{\mathrm{ev}}$ therefore define our smearing 
scheme. We maintain an approximately constant physical link-smearing radius 
$(r_{\mathrm{link}}/a)^2 = \rho n_{\rho}$ by tuning $n_{\rho}$ appropriately. The quark smearing procedure is defined by 
retaining all eigenvectors with eigenvalue $\lambda \lesssim (a\sigma_{\rm s})^2$, where 
$\sigma_{\rm s} = 1\mathrm{GeV}$. As the physical volume ($V$) is increased the 
number of eigenvectors must be scaled as $N_{\mathrm{ev}} \propto V$. 
The stout smearing parameters and 
$N_{\mathrm{ev}}$ are given in Tab.~\ref{t:laph}.   

We employ interpolating operators with light quarks only, 
since we calculate elastic pion-pion scattering amplitudes. The number 
of required light quark Dirac matrix inversions per configuration, denoted $N_{\rm D}$, is also 
given in Tab.~\ref{t:laph}. 
Our treatment of all-to-all propagators enables us to efficiently evaluate all 
required Wick contractions involving two-pion and single-$\rho$ interpolators, which are enumerated in Ref.~\cite{Morningstar:2011ka}. An unbiased estimator 
results only if each quark line in a diagram employs independent stochastic 
sources.  As discussed in 
Ref.~\cite{Bulava:2017stw}, in each diagram we typically average over some 
number of 
multiple noise `orderings', i.e. different permutations of the $N_{\rm R}$ available 
quark lines.   

The correlation functions used in pion-pion scattering require smeared 
quark fields only. However, correlation functions     
for the timelike pion form factor contain the unsmeared vector current operator. These current correlation functions 
are easily constructed in the stochastic LapH framework although they 
employ quark fields which are not projected onto the LapH subspace. 

As done in Ref.~\cite{Mastropas:2014fsa}, by exploiting $\gamma_5$-hermiticity it can be ensured that quark fields 
in the vector current bilinear are always unsmeared sinks 
$\phi_{rb} = Q\,\varrho_{rb}$. 
This motivates the construction of `current sinks' defined as
\begin{align}\label{e:curr}
	J^{(\boldsymbol{d}, \Lambda)}_{rb;r'b'}(t) = 
	\sum_{\textbf{x},\textbf{y}} \phi_{rb}^{\dagger}(x) \, 
	\Gamma^{(\boldsymbol{d}, \Lambda)}(\textbf{x},\textbf{y}) \, \phi_{r'b'}(y),    
\end{align}
where $t=x_0=y_0$ and $\Gamma^{(\boldsymbol{d}, \Lambda)}$ denotes projection onto an irreducible
representation (irrep)  $\Lambda$ of the little group of total momentum 
$\boldsymbol{d}$. $J^{(\boldsymbol{d}, \Lambda)}(t)$ has two 
noise/dilution indices and can therefore be 
employed in the correlation 
function construction procedure of Ref.~\cite{Morningstar:2011ka} exactly as a 
smeared $\rho$-meson interpolator. 

As suggested by Eq.~\ref{e:laph}, in order to save disk space the 
quark sinks are typically projected onto the LapH subspace before they are 
written to disk. However, the  
current functions of Eq.~\ref{e:curr} must be constructed from
unprojected sinks. For fixed quark lines the $\{\phi_{rb}\}$ must 
be kept in memory until calculation of $J^{(\boldsymbol{d}, \Lambda)}(t)$ is 
complete. After construction of the current functions, the quark sinks are 
smeared and written to disk in the usual way. 

The calculation of the Laplacian eigenvectors is performed using a variant of 
the thick restarted Lanczos method~\cite{WuSimon}, which entails global 
re-orthogonalizations of the Krylov subspace. These re-orthogonalizations 
scale poorly with $N_{\mathrm{ev}}$, so that as $L$ is increased calculation of the 
Laplacian eigenvectors will eventually dominate the computational cost. 
However, for the $L\lesssim 5.5\,\mathrm{fm}$ volumes considered here the Dirac matrix inversions are still most computationally intensive.  

We perform these Dirac matrix inversions using the efficient \texttt{DFL\_SAP\_GCR} solver in the \texttt{openQCD} software suite.\footnote{\url{http://luscher.web.cern.ch/luscher/openQCD/}} 
In summation, our workflow consists of three main tasks: (1) Dirac matrix 
inversion, (2) hadron source/sink construction, and (3) formation of correlation functions.
Due to their large storage footprint, the Laplacian eigenvectors are computed 
first in task 1 and not saved to disk. They are then recomputed during task 2, which is implemented entirely in 3-D. Task 3 then no longer requires any
lattice-wide objects and is simply tensor contraction of noise-dilution 
indices.

These different tasks are typically performed on different 
computer architectures, but a rough breakdown of the relative cost is 
$70-80\%$ for the Dirac matrix inversions, $20-26\%$ for task 2, and $1-5\%$ 
for task 3. In total, $1-3\%$ of the total for these three tasks is spent on 
calculating Laplacian eigenvectors. 

\subsection{Finite-volume energies and matrix elements}\label{s:ana}

We consider all elastic energy levels in isovector irreps where the 
$J^{PG}=1^{-+}$ partial wave is the leading contribution up to total momentum 
$\boldsymbol{d}^2 \le 4$, which are tabulated in Tab.~\ref{t:irr}. To 
calculate the energies, we follow the procedure of Refs.~\cite{Bulava:2016mks,Brett:2018jqw}, which is outlined below. 

Outside the resonance region, interacting finite-volume two-pion energies 
are close to their non-interacting values, while for levels with $E_{\cm}$ near $m_{\rho}$ 
these gaps are larger. To exploit the small differences outside 
the resonance region and to treat all energies in a unified manner, 
we employ the ratio fits described in Ref.~\cite{Bulava:2016mks}. 
Using this method, we construct ratios 
\begin{align}\label{e:rat}
	R_n(t) &= \frac{\hat{C}_{n}(t)}{C_{\pi}(\boldsymbol{p}_{1}^2, t) \, 
	C_{\pi}(\boldsymbol{p}_{2}^{2},t)}, 
\\\nonumber	
	\hat{C}_n &= (v_n(t_0,t_{\rm d}), C(t) v_n(t_0, t_{\rm d})) 
\end{align}
where $(\boldsymbol{p}_{1}, \boldsymbol{p}_{2})$ are  momenta of the 
constituent pions in 
the nearest 
non-interacting level and $C_{\pi}(\boldsymbol{p}^2,t)$ is a single-pion correlation 
function with momentum $\boldsymbol{p}^2$. The vector $v_n(t_0,t_{\rm d})$ is a 
generalized eigenvector of the correlator matrix $C(t)$ solving the 
generalized eigenvalue problem (GEVP) 
$C(t_{\rm d}) v_n = \lambda_n C(t_0) v_n$~\cite{Michael:1985ne,Luscher:1990ck}.
The $\{v_n\}$ are used to define the diagonal correlators 
$\hat{C}_n(t)$ between operators with optimal overlap onto the $n$th level, 
and are determined for a single $(t_0,t_{\rm d})$ only. The fitted energies 
vary little as these diagonalization times, as well as the operator basis, are 
varied. 

The difference $\Delta E_n$ between an energy and its closest 
non-interacting $\pi\pi$ counterpart is extracted directly using single-exponential fits to the ratio in Eq.~\ref{e:rat}. Alternatively, the 
interacting energy may be obtained from single- or two-exponential fits to 
$\hat{C}_n$ directly. 
All of these correlated-$\chi^2$ fits are performed over some time range 
$[\tmin,\tmax]$, the variation of which should not affect the fitted 
energies for asymptotically large $t$. Energies obtained from ratio, single-, and two-exponential fits 
all typically depend little on $\tmax$, while ratio fits typically exhibit 
a reduced dependence on $\tmin$ as well. However, the excited state 
contamination in ratio fits may be non-monotonic leading to `bumps' in $\tmin$ 
plots. As an important consistency check, we check agreement of the energies obtained from 
these three types of fits, different $(t_0,t_{\rm d})$ combinations, and 
GEVP operator sets.  

Our fit ranges are chosen conservatively so that the systematic errors 
discussed above due to the GEVP and fit ranges are smaller than the 
statistical ones. This is demonstrated using extensive comparisons of 
$\tmin$-plots for different fit types, $(t_0,t_{\rm d})$ choices, and
GEVP operator bases, similar to Refs.~\cite{Bulava:2016mks,Brett:2018jqw}. 
Bootstrap resamples of all reweighted correlation functions are publicly 
available in 
HDF5 format\footnote{\url{https://doi.org/10.5281/zenodo.1341045}}, as is a python Juypter 
notebook\footnote{\url{https://github.com/ebatz/jupan}} which performs the entire analysis chain. This tool
not only provides an interface to view systematics related to our choices of fitting procedure, 
fitting ranges, and GEVP, but also enables direct access to all results  at each step.  
Generally, in physical units we take $(t_0,t_{\rm d}) \approx (0.5,0.9)\mathrm{fm}$, 
$\tmin = 0.7-1.3\,\mathrm{fm}$, and $\tmax = 2-2.6\,\mathrm{fm}$. 

In addition to determining the energies, on the three finest lattice spacings  
we calculate matrix elements of the electromagnetic current
\begin{align}
	j_{\mu}^{\rm em} = \frac{2}{3}\bar{u}\gamma_{\mu} u - 
	\frac{1}{3} \bar{d} \gamma_{\mu} d  + 
	\dots  
\end{align}
where the ellipsis denotes contributions from heavier quarks. For the
vacuum-to-$\pi\pi$ matrix elements considered in this work, we require insertions of  
the isovector component and a dimension-five counterterm required to implement
$\mathrm{O}(a)$-improvement  
\begin{align}\label{e:ins}
	V_{\mu}^{a} = \bar{\psi} \gamma_{\mu} \frac{\tau^{a}}{2} \psi , 
	\qquad \tilde{\partial}_{\nu} T_{\mu\nu}^{a} = i \tilde{\partial}_{\nu} \bar{\psi} \sigma_{\mu\nu} \frac{\tau^{a}}{2} \psi, 
\end{align}
where $\psi = (u, \, d)^{T}$, $\tau^{a}$ the usual Pauli matrices in 
isospin space, $\sigma_{\mu\nu} = \frac{i}{2}[\gamma_{\mu}, \gamma_{\nu}]$, 
and $\tilde{\partial}_\mu$ the symmetrized lattice derivative. The 
isovector index $a$ is taken to be maximal and henceforth omitted. 

The determination of the timelike pion form-factor then 
employs linear combinations
\begin{align}\label{e:proj}
	V^{(\Lambda, \boldsymbol{d})} = \sum_{\mu} b_{\mu}^{(\Lambda,\boldsymbol{d})}
	V_{R,\mu}, \qquad \sum_{\mu} b_{\mu}^{(\Lambda,\boldsymbol{d})*}
	\,  b_{\mu}^{(\Lambda,\boldsymbol{d})} = 1 
\end{align}
where the coefficients $b_{\mu}^{(\Lambda,\boldsymbol{d})}$ project the current
onto (a row of) irrep $\Lambda$ and spatial momentum $\boldsymbol{d}$. 
The vector current bilinear appearing in Eq.~\ref{e:proj} has been renormalized
and $\mathrm{O}(a)$-improved non-perturbatively according to 
\begin{align}\label{e:ren}
	(V_{{\rm R}})_{\mu} = \zv \left( 1
	+ a\bv \, m_{\rm l} + a\overline{b}_{\rm V}\, \mathrm{tr} \, M_{q}  \right) (V_{\rm I})_\mu, 
	\qquad 
	(V_{\rm I})_\mu = V_{\mu} + a\cv \tilde{\partial}_{\nu} T_{\mu\nu}, 
\end{align}
where the renormalization and improvement coefficients 
$\zv$, $\bv$, $\overline{b}_{\rm V}$, and $\cv$ are functions of the 
 gauge coupling only in this mass-independent scheme. 

We take $\tilde{Z}_{\rm V} =  \zv \left( 1
	+ a\bv \, m_{\rm l} + a\overline{b}_{\rm V}\, \mathrm{tr} \, M_{q} \right)$ and 
	$\cv$ from the non-perturbative 
	determination of Ref.~\cite{agcv}. 
An alternative determination of the non-singlet current renormalization 
constants for this
lattice discretization is found in Ref.~\cite{DallaBrida:2018tpn}.  
Another preliminary non-perturbative
determination of $\zv$ can be found in Ref.~\cite{Heitger:2017njs}, while 
non-perturbative determinations of $\bv$ and $\overline{b}_{\rm V}$
 are performed in Refs.~\cite{Korcyl:2016ugy,Fritzsch:2018zym}.  

Operationally, we calculate current correlation functions using 
Eq.~\ref{e:curr} for both the dimension four and five operators in 
Eq.~\ref{e:ins} projected onto definite momentum $\boldsymbol{d}$ and 
irrep $\Lambda$. These current correlation functions, which are vectors in the GEVP index, are given as  
\begin{align}\label{e:ccor}
	D^{(\Lambda,\boldsymbol{d})}(t-t_0) = \langle J^{(\Lambda,\boldsymbol{d})}(t) 
	\bar{O}^{(\Lambda,\boldsymbol{d})}(t_0) \rangle, 
\end{align}
where $J$ denotes either of the operators in Eq.~\ref{e:ins} projected according to Eq.~\ref{e:proj} and 
$\hat{O}^{(\Lambda,\boldsymbol{d})}$ is an interpolator for irrep 
$(\Lambda,\boldsymbol{d})$.

To extract the finite volume matrix elements 
$\langle 0 | \hat{V}^{(\Lambda,\boldsymbol{d})} | \Lambda \boldsymbol{d} n 
\rangle$, we calculate the current correlation functions (defined in Eq.~\ref{e:ccor}) containing each the two 
operators in Eq.~\ref{e:ins}. 
These are used to form optimized current correlation functions using the GEVP eigenvectors
\begin{align}
	\hat{D}_n(t) = \left( D(t), v_n \right), 
\end{align}
where the inner product is taken over the GEVP index. 
Using these optimized current correlators, we then construct three 
ratios which plateau to the desired matrix elements asymptotically for
large $t$ (up to GEVP systematics)
\begin{align}\label{e:curat}
	R^{(1)}_n(t) &= \left|\frac{\hat{D}_n(t)}{\sqrt{\hat{C}_n(t) \mathrm{e}^{-E_n t}}}\right|,
	\qquad
	R^{(2)}_n(t) &= \left|\frac{\hat{D}_n(t)}{A_n \mathrm{e}^{-E_n t}}\right|,
	\qquad
	R^{(3)}_n(t) &= \left|\frac{\hat{D}_n(t)\, A_n}{\hat{C}_n(t)}\right|,
\end{align}
where $A_n$ and $E_n$ are determined previously from the ratio fits 
to $\hat{C}_n(t)$. The final matrix elements are then obtained from a plateau 
average of these ratios over a range $[\tmin,\tmax]$. 

Each of these ratios possesses different excited state contamination. In 
analogy with the determination of the energies discussed above, consistency of 
the matrix elements using different ratios, $(t_0,t_d)$, and GEVP bases provides
a stringent check in their determination.
GEVP corrections to $A_n$ have a different form than those of the energies~\cite{Blossier:2009kd}, and 
our choices of fit ranges are optimized with the energies in mind. 
For this reason we take  
$R_n^{(1)}(t)$ in Eq.~\ref{e:curat} as the best estimate of the matrix elements. 
Nonetheless, all three ratios are typically consistent.  
Data illustrating the 
$\tmin$-dependence of these ratios and the comparisons mentioned above may
also be found in the HDF5 files and Jupyter notebook.
After determining the matrix elements for each current operator, we combine them to form the renormalized combination in 
Eq.~\ref{e:ren}. 

\subsection{Amplitudes from finite volume energies and matrix elements}\label{s:form} 

First we determine the elastic $I=1$ $p$-wave pion-pion 
scattering amplitude using the finite-volume energies discussed in 
Sec.~\ref{s:ana}. This amplitude is obtained from the determinant condition 
introduced for general two-to-two scattering in Eq.~\ref{e:det}. However, 
considerable simplification occurs for elastic scattering between spinless 
identical particles. Here the $K$-matrix is diagonal in orbital angular 
momentum $\ell$ with trivial structure in the irrep occurrence index $n_{\rm occ}$. The 
box matrix $B$, which encodes the effect of the finite periodic 
spatial volume, mixes different orbital angular momenta and is dense in $n_{\rm occ}$. 

All the irreps used in this work are given in Tab.~\ref{t:irr} together with 
the pattern of partial wave mixing induced by the infinite-dimensional 
$B$-matrix. Explicit expressions for all $B$-matrix elements up to 
$\ell \le 6$ are given in Ref.~\cite{Morningstar:2017spu}.
\begin{table}
\centering
\begin{tabular}{c c l}
  \toprule
	$\dvec$ & \textbf{$\Lambda$} & $\ell$ \\
	\midrule
	$(0,0,0)$ &$T^{+}_{1u}$& 1, 3, $5^2$,  \ldots \\
\midrule%
	$(0,0,n)$ &$A^{+}_1$&  1, 3, $5^2$,\ldots \\
&$E^{+}$& 1, $3^2$, $5^3$, \ldots \\
\midrule%
	$(0,n,n)$ &$A^{+}_1$& 1, $3^2$, $5^3$, \ldots \\
	&$B^{+}_1$& 1, $3^2$, $5^3$ \ldots \\
	&$B^{+}_2$& 1, $3^2$, $5^3$, \ldots \\
\midrule%
	$(n,n,n)$ &$A^{+}_1$& 1, $3^2$, $5^2$, \ldots \\
	&$E^{+}$& 1, $3^2$, $5^4$, \ldots \\
\bottomrule
\end{tabular}
	\caption{\label{t:irr} Finite-volume irreps $\Lambda$ (second column) of the little group for various classes of  
	total momenta $\boldsymbol{P}_{\rm tot} = (2\pi/L)\dvec$ (first column) 
	employed here. The superscripts on the partial waves $(\ell)$ contributing to 
	that irrep denote the number of multiple occurrences, while the `+' indicates positive $G$-parity.}
\end{table}
Several off-diagonal $B$-matrix elements vanish for identical particles, 
preventing partial wave mixing between even and odd $\ell$. If contributions 
from $\ell \ge 3$ partial waves are neglected, this simplification provides a 
one-to-one correspondence between energies in the irreps of Tab.~\ref{t:irr} 
and 
\begin{align}\label{e:kmat}
	\tilde{K}^{-1}_{11}(E_{\cm}) = \left(\frac{q_{\cm}}{m_{\pi}}\right)^3 \cot \delta_{1}(E_{\cm}), 
\end{align}
where $q_{\cm}$ is the center of mass momentum and $\delta_1$ the $I=1$ $\pi\pi$ phase shift.
This approximation is justified by the near-threshold suppression of higher 
partial waves and to test it we perform global fits including $f$-wave 
contributions. Refs.~\cite{Morningstar:2017spu,Wilson:2015dqa} perform 
similar fits and find such contributions negligible. 

We turn now to the determination of the timelike pion form factor 
$|F_{\pi}(E_{\cm})|$ using the finite-volume matrix elements calculated 
according to Sec.~\ref{s:form}. The relations employed here for zero-to-two 
matrix elements are given in 
Refs.~\cite{Meyer:2011um,Feng:2014gba,Briceno:2015csa} which are based on the 
seminal work of Ref.~\cite{Lellouch:2000pv}.  

We first define the angle $\phi^{(\boldsymbol{d},\Lambda)}_1$ using $B^{(\boldsymbol{d},\Lambda)}_{11} = 
(q_{\cm}/m_{\pi})^3\cot{\phi^{(\boldsymbol{d},\Lambda)}_1}$ where $B$ is from Eq.~\ref{e:det}. This 
pseudophase is used together with the physical phase shift to relate the finite- and infinite-volume matrix elements  
\begin{align}\label{e:llm}
	\left| F_{\pi}(E_{\cm})\right|^2 = g_{\Lambda}(\gamma) \, 
	\left( q_{\cm} \frac{\partial \delta_1}{\partial q_{\cm}} + 
	u \frac{\partial \phi_1^{(\boldsymbol{d},\Lambda)}}{\partial u} \right)
	\, \frac{3\pi E_{\cm}^2}{2q_{\cm}^5 L^3} \, \left|\langle 0 | 
	V^{(\boldsymbol{d},\Lambda)}|\boldsymbol{d}\Lambda n\rangle \right|^2, 
\end{align}
where $q_{\cm}$ is the magnitude of the center-of-mass (three) momentum, 
$u^2 = L^2q_{\cm}^2/(2\pi)^2$, and 
\begin{align}
	g_{\Lambda}(\gamma) = \begin{cases} 
		\gamma^{-1}, &  \Lambda = A_{1}^{+} \\
		\gamma, & \mathrm{otherwise}
	\end{cases}
\end{align}
where $\gamma = E/E_{\cm}$. The infinite volume matrix elements are therefore 
obtained from their finite volume counterparts using the multiplicative 
Lellouch-L\"{u}scher-Meyer (LLM) 
factor shown on the r.h.s of Eq.~\ref{e:llm}. 

As is evident from Eq.~\ref{e:llm}, determination of $|F_{\pi}(E_{\cm})|$ 
requires not only the finite-volume matrix element $\left|\langle 0 | 
	V^{(\boldsymbol{d},\Lambda)}|\boldsymbol{d}\Lambda n\rangle \right|$ but 
	also the derivative of $\delta_1$. This derivative is obtained from a 
	parametrization of the phase shift points described above and covariances 
	between all data are treated explicitly using the bootstrap procedure. 
	Parametrization of $\delta_1(E_{\cm})$ and $|F_{\pi}(E_{\cm})|$ is discussed 
	in Sec.~\ref{s:res}.

%% file: results.tex
\section{Results}\label{s:res} 

We first present results for the elastic $\pi\pi$ scattering amplitude. As 
discussed in Sec.~\ref{s:form}, if $\ell \ge 3$ contributions to Eq.~\ref{e:det}
are neglected there is a one-to-one correspondence between finite-volume 
energies and $\tilde{K}_{11}(E_{\cm})$ defined in Eq.~\ref{e:kmat}. 
This energy dependence is parametrized by a Breit-Wigner shape
\begin{align}\label{e:bw}
	\tilde{K}_{11}^{-1}(E_{\cm}) = 
	\left(\frac{m^2_{\rho}}{m^2_{\pi}} - \frac{E^2_{\cm}}{m_{\pi}^2}\right)
	\frac{6\pi E_{\cm}}{g_{\rho\pi\pi}^2m_{\pi}}
\end{align}
involving two free parameters $g_{\rho\pi\pi}^2$ and $m^2_{\rho}/m^2_{\pi}$. 
A correlated-$\chi^2$ fit of all points is performed according to 
the `determinant residual' method of 
Ref.~\cite{Morningstar:2017spu} with $\mu = 10$, although without $\ell\ge 3$ contributions 
the $\tilde{K}^{-1}$- and $B$-matrices are one-dimensional so that the determinant is 
trivial.  Results for these fit parameters,  which 
are both constrained to be positive, as well 
as the $\chi^2$ per degree of freedom, $\hat{\chi}^2 = \chi^2/N_{\rm d.o.f}$, are given in Tab.~\ref{t:pres} for each of the 
ensembles employed here. 
\begin{table}
\centering
\renewcommand*{\arraystretch}{1.1}
	\input{table_bw.tex}	
	\caption{\label{t:pres} Results of correlated-$\chi^2$ fits of the 
	$I=1$ elastic $p$-wave $\pi\pi$ amplitude to Eq.~\ref{e:bw}. After the 
	ensemble ID and the number of levels, the three subsequent columns show results from fits ignoring $\ell=3$ contributions. The remaining columns contain results from fits including the $f$-wave contribution as described in the text.} 
\end{table}

The influence of $\ell \ge 3$ partial waves is assessed  
by enlarging the determinant condition of 
Eq.~\ref{e:det} to include the $\ell=3$ contributions noted in 
Tab.~\ref{t:irr}. For this fit the $f$-wave is parametrized by an 
unconstrained constant $\tilde{K}_{33}^{-1}(E_{\cm}) = -(m_{\pi}^7a_3)^{-1}$ 
yielding results which are also shown in Tab.~\ref{t:pres}.
We see therefore that there is little dependence on including $\ell\ge 3$. 

\begin{figure}
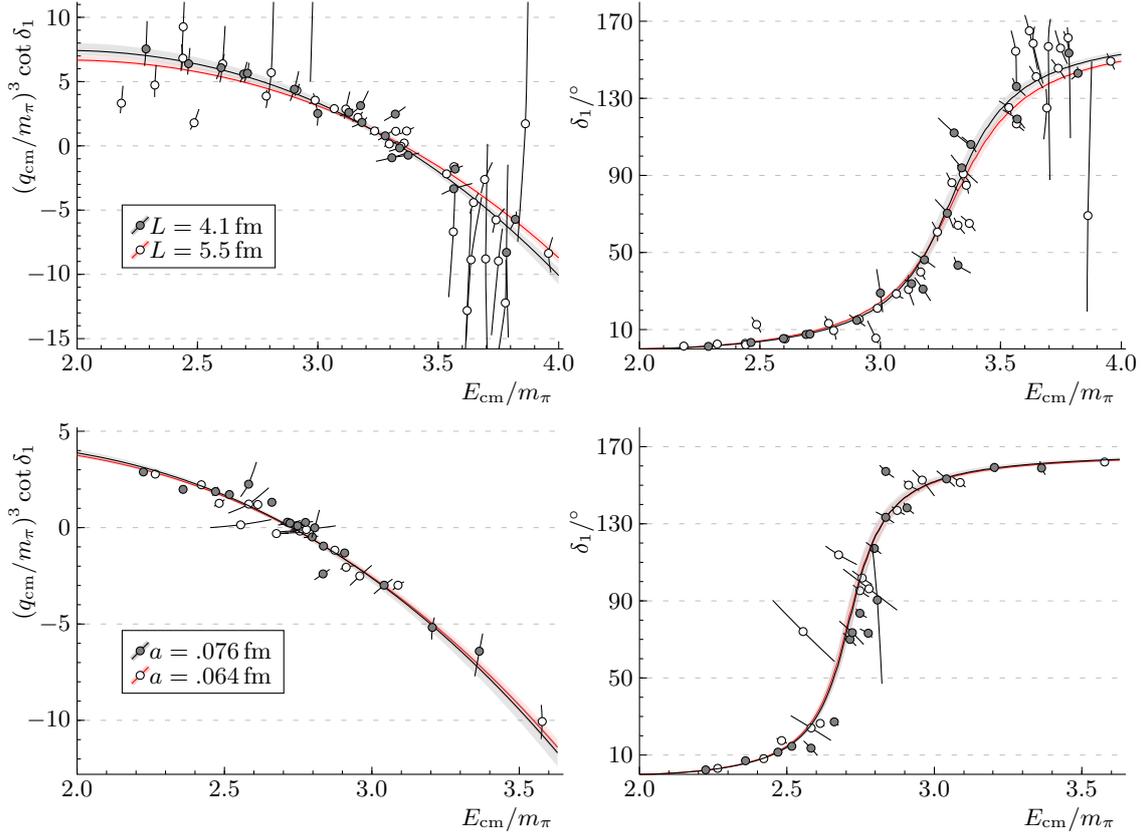

	\includegraphics{tikz-figures/combiPlot_d101_c101.tikz}
	\includegraphics{tikz-figures/combiPlot_n200_n401.tikz}
	\caption{\label{f:comp}\emph{Top row}: Comparison of $\tilde{K}^{-1}_{11}(E_{\cm})$ (left) and $\delta_1(E_{\cm})$ (right) between the C101 and D101
	ensembles, which have the same parameters but different physical volumes.
	\emph{Bottom row}: The same comparison for the $N200$ and $N401$ ensembles (white and gray markers respectively), 
	which have (approximately) the same quark masses but different lattice 
	spacings.} 
\end{figure}

Interested readers may perform further fits using App.~\ref{a:box}, where 
energies and phase shift points for all ensembles (neglecting $\ell \ge 3$) 
are tabulated and plotted, or the Jupyter notebook described in Sec.~\ref{s:ana} where bootstrap samples of all energy levels are available. However, it is worth comparing some of the ensembles in this 
data set here. An explicit check of finite volume effects using the C101 
and D101 ensembles, which have the same parameters but different 
volumes, is shown in Fig.~\ref{f:comp}. 
That figure also shows a comparison between the N401 and N200 ensembles, which 
have (approximately) the same quark masses but different lattice spacing. 
It is thus evident on these ensembles that both effects are not visible within 
our statistical errors. Finally, all results for $m_{\rho}$ and $g_{\rho\pi\pi}$
in shown in Fig.~\ref{f:pcmp}, where they are converted to physical units
using the scale determined in Ref.~\cite{Bruno:2016plf}. 

\begin{figure}
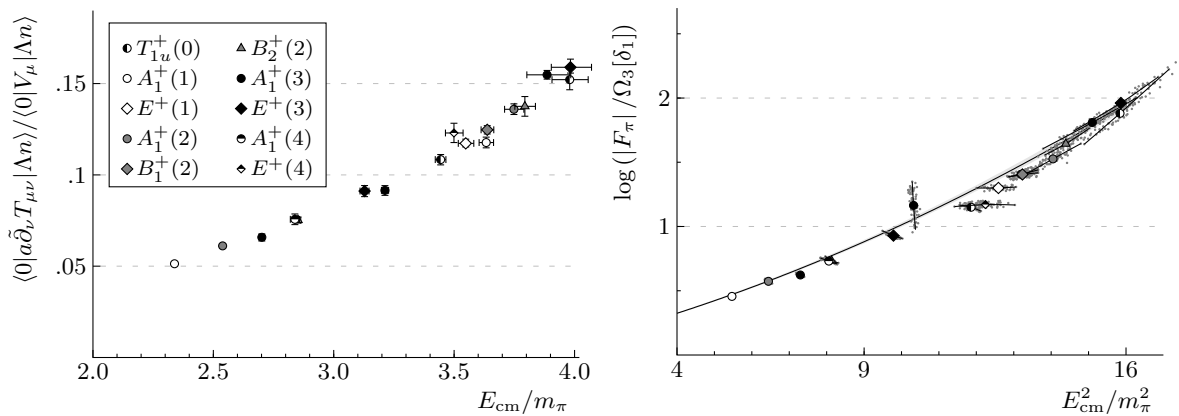

	\includegraphics{tikz-figures/d200_oacuplot.tikz}
	\includegraphics{tikz-figures/d200_dispFit_nSubtr3.tikz}
	\caption{\label{f:fcomp}\emph{Left}: ratio of matrix 
	elements for the $\mathrm{O}(a)$ counterterm over the leading matrix element 
	for the D200 ensemble.
	\emph{Right}: Thrice-subtracted dispersive fit to $\ln \,Q_3(s)$ on the D200 ensemble. The central 68\% of bootstrap samples (thinned out for clarity) are 
	shown for each individual point, together with the best-fit line.} 
\end{figure}

We turn now to results for the $I=1$ timelike pion form factor, which are 
determined according to Sec.~\ref{s:ana}.  
Results for the form factor from  
 Eq.~\ref{e:llm} are also tabulated in App.~\ref{a:box}. As discussed in
 Sec.~\ref{s:ana}, 
 we employ a non-perturbative 
determination~\cite{agcv} of $\cv$ multiplying the dimension-five counter term 
in 
Eq.~\ref{e:ren}. Apart from the one from Ref.~\cite{agcv}, there is the 
preliminary determination of Ref.~\cite{Heitger:2017njs} which obtains values 
larger in magnitude
using a different improvement condition. If $\cv$ is non-negligible, the 
relative 
magnitude of the leading order matrix elements to this counterterm 
is of interest. Their ratio is shown in Fig.~\ref{f:fcomp}. 
Given its $5-15\%$ size, a larger $\cv$ could indicate 
larger cutoff effects in the form factor than we observe using $\cv(g_0)$ from 
Ref.~\cite{agcv}, which is at the few-percent level.

We now turn to parametrization of the form factor. Ref.~\cite{Feng:2014gba}
employs the Gounaris-Sakurai parametrization~\cite{Gounaris:1968mw} 
\begin{gather}\label{e:gs}
	F_{\pi}^{GS}(\sqrt{s}) = \frac{f_0}{q_{\rm cm}^2h(\sqrt{s}) - q_{\rho}^2 h(m_{\rho})
		+ b(q_{\rm cm}^2 - q_{\rho}^2) - \frac{q_{\rm cm}^3}{\sqrt{s}}i},
		\\\nonumber 	
		b = -h(m_{\rho}) - \frac{24\pi}{g_{\rho\pi\pi}^2} - \frac{2q_{\rho}^2}{m_{\rho}}h'(m_{\rho}), \qquad f_{0} = -\frac{m_{\pi}^2}{\pi} - q_{\rho}^2h(m_{\rho}) - b\frac{m^{2}_{\rho}}{4}, \\\nonumber		
		h(\sqrt{s}) =  \frac{2}{\pi} 
		\frac{q_{\rm cm}}{\sqrt{s}} \ln\left(\frac{\sqrt{s} + 2q_{\rm cm}}{2m_{\pi}}\right),
\end{gather}
 where the notation is from Ref.~\cite{Francis:2013qna} and $q_{\rho}$ is 
 the center-of-mass momentum at the resonance energy. This parametrization
 depends only on $m_{\rho}$ and $g_{\rho\pi\pi}$, and therefore 
 describes the form factor with no additional free parameters.

 Additional parametrizations are suggested by unitarity constraints. In the 
 elastic approximation, the form factor satisfies the $n$-subtracted dispersion 
 relation
 \begin{align}
	 F_{\pi}(s) = \sum_{k=0}^{n-1} \frac{s^k}{k!}\frac{d^{k}}{ds^k}F_{\pi}(0) + 
	 \frac{s^n}{\pi} \int_{4m_{\pi}^2}^{\infty} \frac{dz}{z^n} \frac{\tan \delta_1(z) \mathrm{Re}\, F_{\pi}(z) }{z - s - i\epsilon}.
 \end{align}
This dispersion relation has the Omn\`{e}s-Muskhelishvili
solution~\cite{Omnes:1958hv,musk} of
\begin{align}\label{e:omf}
	F_{\pi}(s) &= Q_n(s) \, \exp \left\{ \frac{s^n}{\pi}\int_{4m_{\pi}^2}^{\infty}
	\frac{dz}{z^n} \frac{\delta_1(z)}{z -s -i\epsilon}\right\}
\\\nonumber
	& = Q_n(s) \, \Omega_n[\delta_1](s), \qquad\quad \ln  \, Q_n(s) = \sum_{k=1}^{n-1} p_k \, s^k,
\end{align}
where due to charge conservation the $k=0$ term vanishes, and we have defined the Omn\`{e}s function $\Omega_n[\delta_1](s)$. The constants $p_k$ are fit
parameters and proportional to logarithmic derivatives of $F_{\pi}$. Given the 
Breit-Wigner
parametrization for the phase shift, the twice-subtracted dispersion relation ($n=2$) has a single additional parameter ($p_1$) and the thrice-subtracted ($n=3$) 
two additional parameters, $p_1$ and $p_2$. These parameters appear in 
$\ln \, Q_n(s)$, while
$\Omega_n[\delta_1](s)$ depends only 
	on the Breit-Wigner parametrization of $\delta_1$.  

\begin{figure}
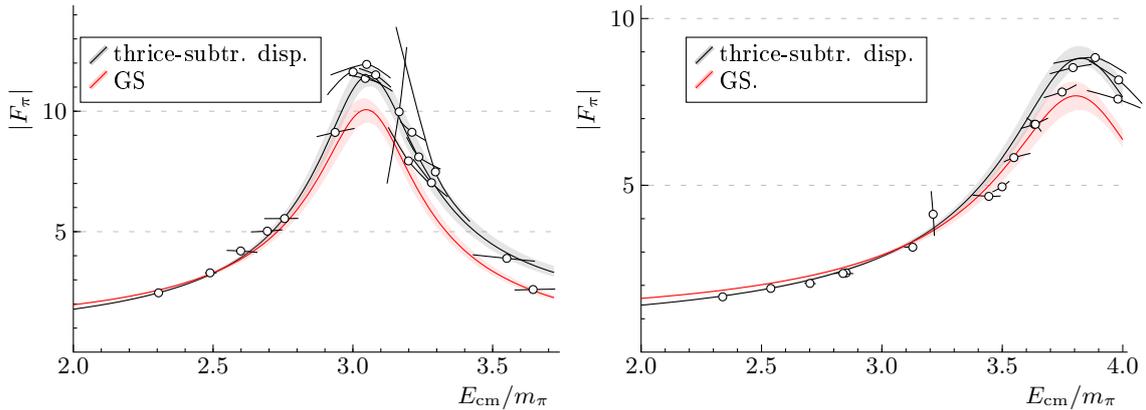

	\includegraphics{tikz-figures/j303_ffPlot_combiPlot.tikz}
	\includegraphics{tikz-figures/d200_ffPlot_combiPlot_gs_disp.tikz}
	\caption{\label{f:gcmp}Form factor results
	shown with both the Gounaris-Sakurai (GS) parametrization and the thrice-subtracted dispersive fit on the J303 ensemble (\emph{left}) and D200 ensemble (\emph{right}). }
\end{figure}

\begin{figure}
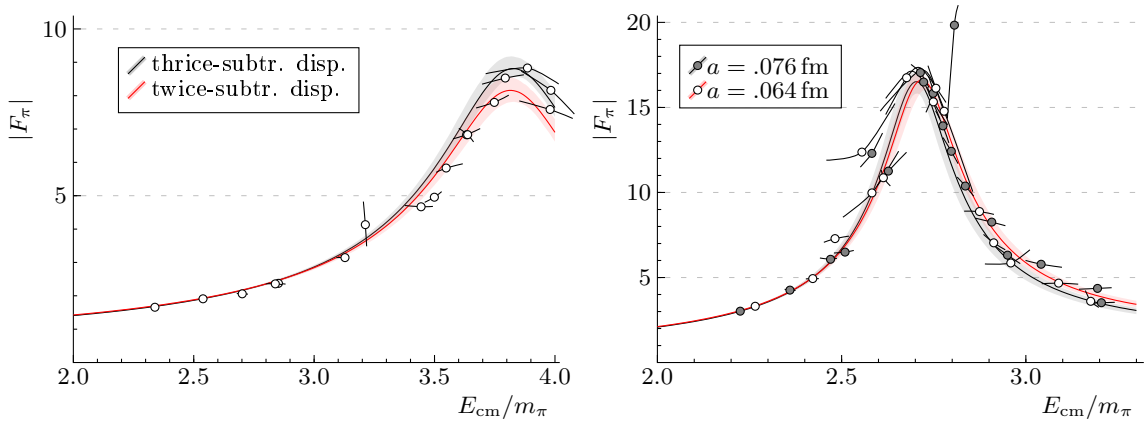

	\includegraphics{tikz-figures/d200_ffPlot_combiPlot.tikz}
	\includegraphics{tikz-figures/combiFFPlot_n200_n401.tikz}
	\caption{\label{f:fcont}\emph{Left}: The twice- and thrice-subtracted dispersive fits on the D200 ensemble. 
	\emph{Right}: Comparison of form factor with thrice-subtracted dispersive fits on
	both the N401 and N200 ensembles, which have similar quark masses and 
	different lattice spacings.}
\end{figure}

For these fits, we isolate $\ln \, Q_n(s)$ by constructing $\ln ( |F_{\pi}| / \Omega_n[\delta_1] )$ and fit it to the appropriate polynomial. An example of a thrice-subtracted fit 
is shown in Fig.~\ref{f:fcomp}, while the twice- and thrice subtracted 
fits are compared in Fig.~\ref{f:fcont} for the D200 ensemble. Finally, the 
Gounaris-Sakurai parametrization is compared to the thrice-subtracted fit 
on the J303 and D200 ensembles in Fig.~\ref{f:gcmp}.

The results from these fits are compared in Tab.~\ref{t:fres}. The large 
$\hat{\chi}^2$ is due to the significant 
correlation between the horizontal and vertical errors, which is visible in 
Fig.~\ref{f:fcomp}. Fits with four subtractions ($n=4$) do not significantly reduce $\hat{\chi}^2$. Results from the N401 and N200 ensembles, which have similar quark masses but 
different lattice spacings, are shown in Fig.~\ref{f:fcont} together with thrice-subtracted fits. 
Agreement between these two ensembles indicates that cutoff effects are also
under control in the form factor.   
\begin{table}
\centering
\renewcommand*{\arraystretch}{1.1}
	\input{table_ff.tex}
	\caption{\label{t:fres} Results from twice- and thrice-subtracted 
	dispersive fits (see Eq.~\ref{e:omf}) to the form factor on each ensemble with the three finest 
	lattice spacings. The large $\hat{\chi}^2$ is due to the significant 
	correlations between $E_{\cm}$ and $\ln \, Q_n$.}
\end{table}

%% file: table_bw.tex
\begin{tabular}{ccccccccc}
\toprule
 & & \multicolumn{3}{c}{$\ell=1$ fits} & \multicolumn{4}{c}{$\ell=1, 3$ fits} \\
 \cmidrule(lr){3-5}\cmidrule(lr){6-9}
ID & $N_{\rm lvl}$ & $m_{\rho}/m_{\pi}$  &    $g_{\rho\pi\pi}$    &  $\hat\chi^2$  & $m_{\rho}/m_{\pi}$  &    $g_{\rho\pi\pi}$    & $m_{\pi}^7 a_3 \times 10^3 $  &  $\hat\chi^2$   \\
\midrule
D101 & 43 & 3.366(15) & 6.19(10) & 2.51 & 3.370(15) & 6.23(10) & -0.56(30) & 2.49 \\
C101 & 21 & 3.395(26) & 5.67(17) & 1.07 & 3.399(30) & 5.72(19) & -0.18(26) & 1.11 \\
	N401 & 19 & 2.717(16) & 5.84(12)  &  1.64 &  2.721(16) & 5.88(13) & -2.7(3.0)  & 1.71 \\
	N200 & 15 & 2.733(16) & 5.94(10) &  1.34 &  2.733(16) & 5.94(10) &  0.0(2.9) & 1.45 \\
D200 & 17 & 3.877(34) & 6.16(19) & 0.81 & 3.883(36) & 6.15(20) & -0.61(94) & 0.84 \\
J303 & 18 & 3.089(25) & 6.30(17) & 0.75 & 3.096(25) & 6.32(17) & -4.2(3.6) & 0.73 \\
\bottomrule
\end{tabular}

%% file: table_ff.tex
\begin{tabular}{ccccccc}
\toprule
& & \multicolumn{2}{c}{$n=2$} & \multicolumn{3}{c}{$n=3$} \\
 \cmidrule(lr){3-4}\cmidrule(lr){5-7}
 ID & $N_{\rm lvl}$ & $m_\pi^2 p_1$  &  $\hat\chi^2$  & $m_\pi^2 p_1$  & $m_\pi^4 p_2$  & $\hat\chi^2$   \\
\midrule
N401 & 18 & 0.1372(13) & 8.9 & 0.099(4) & 0.0144(7) & 5.5  \\
N200 & 15 & 0.1405(18)  & 6.0 & 0.104(7) & 0.0143(12) & 3.6 \\
D200 & 17 & 0.0738(17) & 5.7 & 0.0674(16) & 0.0034(2) & 4.1 \\
J303 & 18 & 0.1129(15) & 2.7 & 0.101(4) & 0.0075(6) & 2.1 \\
\bottomrule
\end{tabular}

%% file: concl.tex
\section{Conclusions}\label{s:concl}

This work presents an $N_{\rm f}=2+1$ calculation of the $I=1$ elastic $p$-wave 
$\pi\pi$ 
scattering phase shift and timelike pion form factor which addresses systematic errors due to the finite 
lattice spacing, mixing of higher partial waves, and residual (exponential) 
finite volume effects. 

While we do not perform continuum and chiral extrapolations here, our data can 
be used for future such extrapolations. Chiral extrapolations of lattice 
scattering data using unitarized extensions of chiral effective 
 theory~\cite{Nebreda:2010wv,Bavontaweepanya:2018yds,Hu:2017wli,Bolton:2015psa,
 Liu:2016wxq,Liu:2016uzk,MartinezTorres:2017bdo,Guo:2016zep,Guo:2018kno} 
 have been performed, although to date 
 cutoff effects have not been considered. Nonetheless, it is 
 evident in our data that cutoff effects in both the scattering amplitude (shown in Fig.~\ref{f:comp}) and 
 the timelike pion form factor (Fig.~\ref{f:fcont}) are small with respect to our statistical 
 errors. The coarser lattice spacing in this comparison is $a=0.075\,\mathrm{fm}$. Furthermore, for the scattering amplitude we also check 
 explicitly (in Fig.~\ref{f:comp}) that finite volume effects are also insignificant at our coarsest lattice spacing. The two volumes used here have 
 $m_{\pi}L = 4.6$ and $6.1$.

\begin{figure}
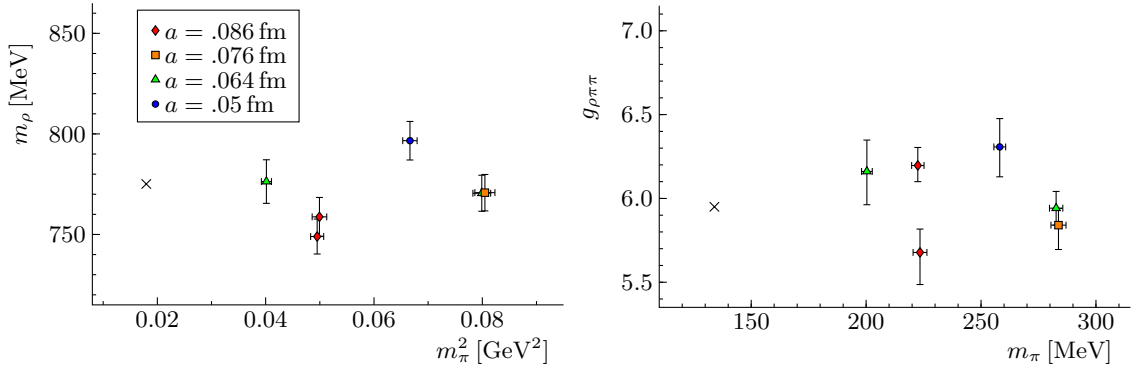

	\includegraphics{tikz-figures/mrSumm.tikz}
	\includegraphics{tikz-figures/gSumm.tikz}
	\caption{\label{f:pcmp} The resonance mass $m_{\rho}$ (\emph{left}) in physical units and the coupling $g_{\rho\pi\pi}$ (\emph{right}) from Breit-Wigner 
	fits to the scattering amplitude on all ensembles.} 
\end{figure}

 As discussed in Sec.~\ref{s:intro}, complete extrapolations of the energy dependence of amplitudes are 
 left for future work. However as a necessary ingredient to determine the 
 form factor, we use Breit-Wigner fits to model the energy dependence of $\delta_1(E_{\cm})$.  
 A summary of the fit results for the resonance mass and coupling are shown in 
Fig.~\ref{f:pcmp}.
 The CLS ensembles employed here adjust 
 $m_{\rm s}$ as $m_{\rm l}$ is lowered to its physical value such that 
 $\mathrm{tr} \, M_{\rm q} = \mathrm{const}.$ is fixed. This is to be contrasted with the more
 common strategy of fixing $m_{\rm s}$ to its physical value for all $m_{\rm l}$ such as  the recent $N_{\rm f} = 2+1$ results in
 Ref.~\cite{Fu:2016itp} which employs rooted staggered fermions. 
Compared with the more standard trajectory, it appears that the slope of $m_{\rho}$ is somewhat flatter here.  
Although not shown in this work, recent summaries of 
existing results for the $\rho$-resonance parameters are found in 
Refs.~\cite{Bulava:2016mks,Alexandrou:2017mpi}. 

In addition to future fits and extrapolations, our results for the timelike 
pion form factor can be used to extend the vector-vector 
correlator as described in Ref.~\cite{Meyer:2018til} and implemented in 
Ref.~\cite{DellaMorte:2017khn} without reliance on 
experiment. This may significantly improve lattice determinations of hadronic
vacuum polarization contribution to anomalous magnetic moment of the muon, 
$a_{\mu}^{\rm HVP}$. 

Finally, the computational effort expended on the CLS lattices for this work
can be largely re-used for other two-to-two amplitude calculations. First work
in this direction for $N\pi$ scattering has already appeared in 
Ref.~\cite{Andersen:2017una}. The set of ensembles used here will also be 
augmented by several others at lighter quark masses, include one with 
$L=6.5\,\mathrm{fm}$ at the physical point, which is presented in Ref.~\cite{Mohler:2017wnb}.

%% file: app.tex
\appendix

\section{Finite volume energies and scattering amplitudes}\label{a:box}

\begin{figure}[ht]
  \centering
  \includegraphics{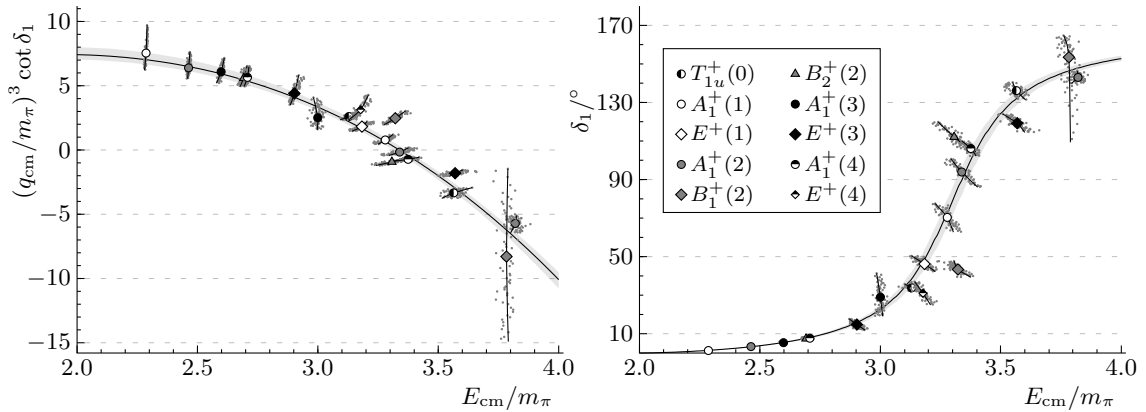}
	\caption{\label{f:c101} $\tilde{K}^{-1}_{11}$ (left) and phase shift (right) on the C101 ensemble, together with the Breit-Wigner fit.}
\end{figure}

\begin{figure}[ht]
  \centering
  \includegraphics{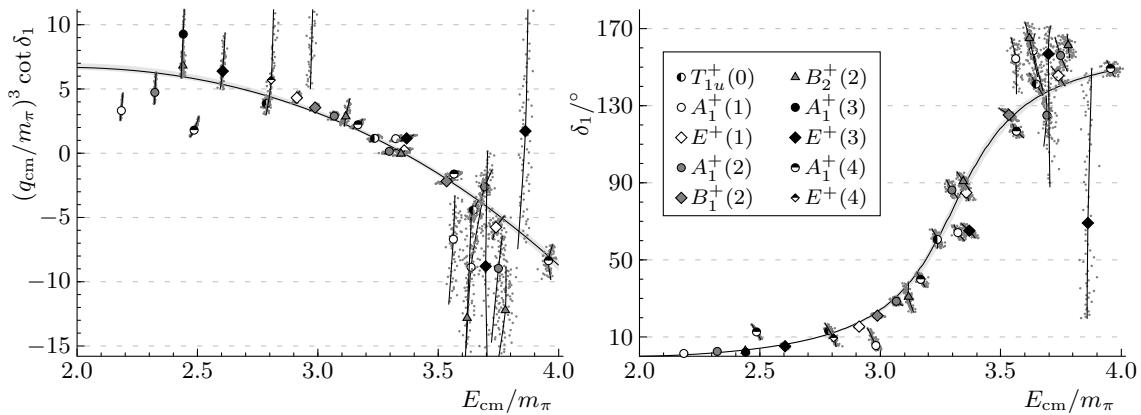}
	\caption{\label{f:d101} Same as Fig.~\ref{f:c101} for the D101 ensemble. 
	One state in each of the $A_1^+(3)$ and $B_2^+(2)$ irreps 
	which have very large errors have been removed from the plot.}
  
\end{figure}

\begin{figure}[ht]
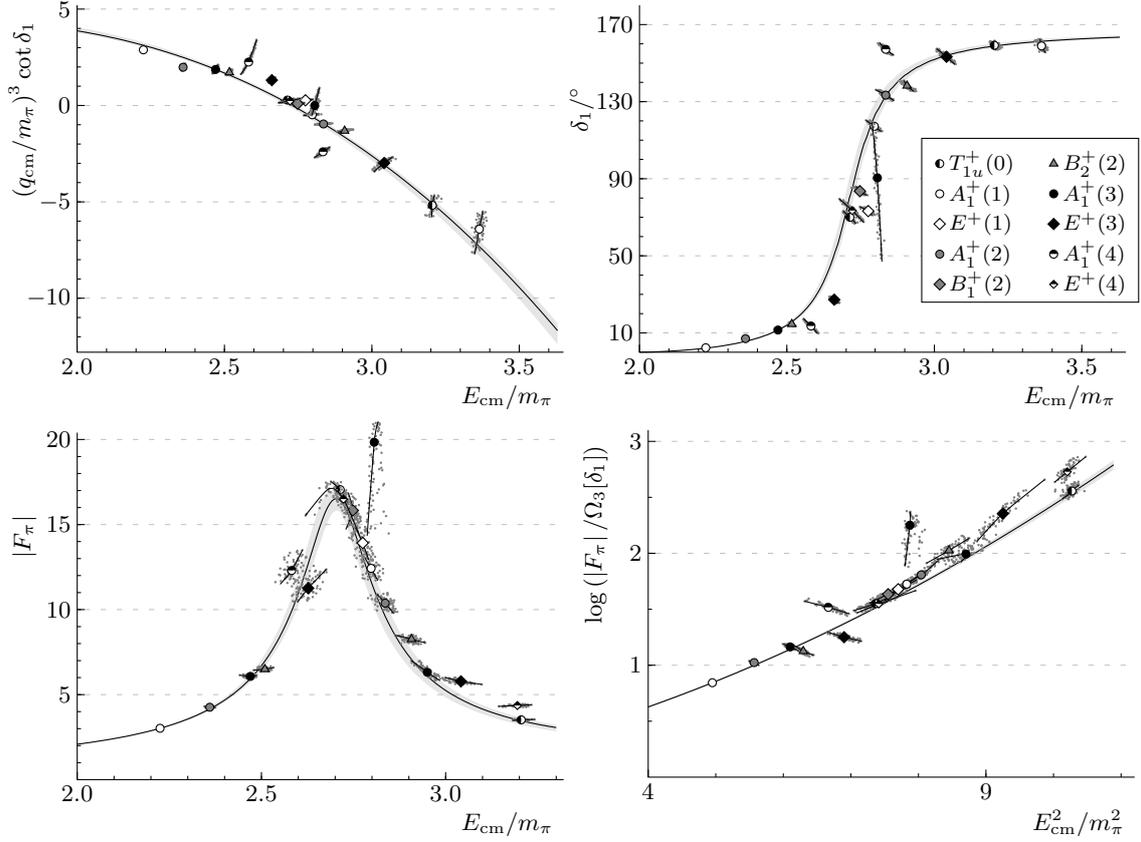

  \centering
	\includegraphics{tikz-figures/n401_ell1plot.tikz}
	\includegraphics{tikz-figures/n401_ffPlot_disp_nSubtr3.tikz}
	\includegraphics{tikz-figures/n401_dispFit_nSubtr3.tikz}
	\caption{\label{f:n401} \emph{Top row}: Same as Fig.~\ref{f:c101} for the N401 ensemble. \emph{Bottom row}: the timelike pion form factor (\emph{left}) and the ratio employed in the thrice-subtracted subtracted dispersive fit (\emph{right}), which is also shown.}
\end{figure}

\begin{figure}[ht]
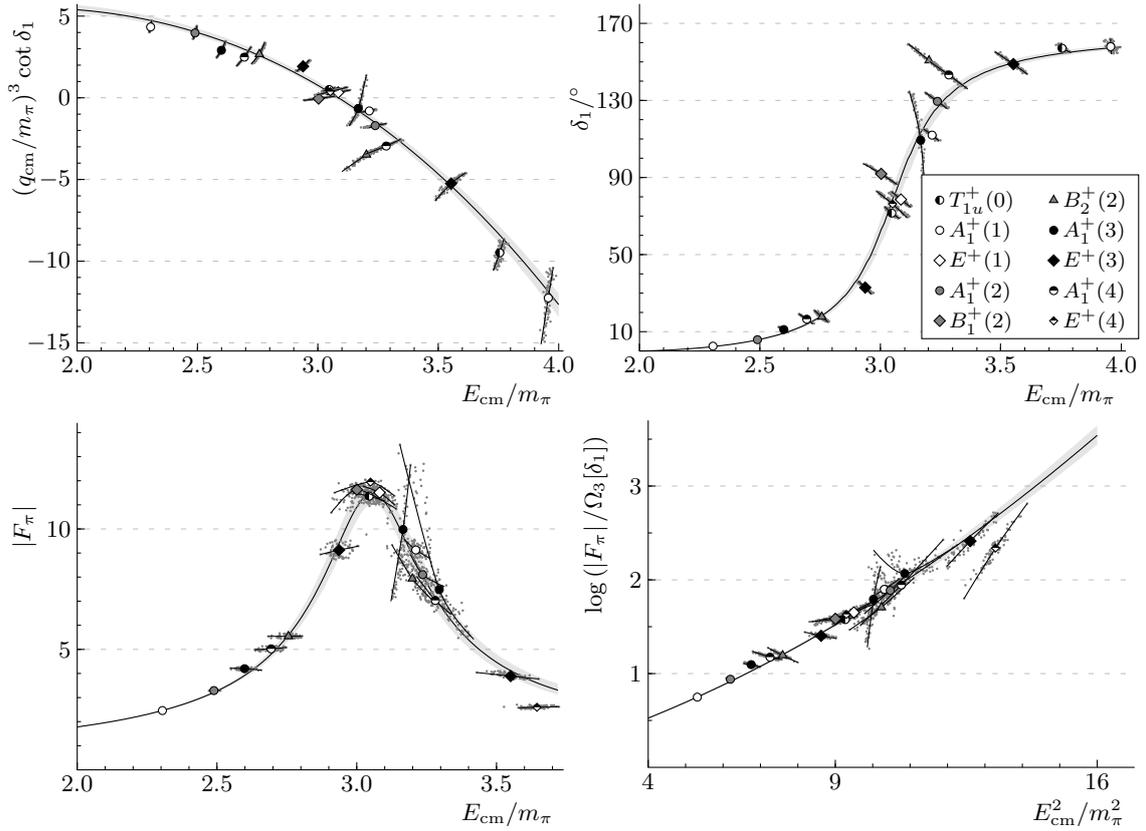

  \centering
	\includegraphics{tikz-figures/j303_ell1plot.tikz}
	\includegraphics{tikz-figures/j303_ffPlot_disp_nSubtr3.tikz}
	\includegraphics{tikz-figures/j303_dispFit_nSubtr3.tikz}
	\caption{\label{f:j303} Same as Fig.~\ref{f:n401} for the J303 ensemble.}
\end{figure}

In this appendix, we tabulate and plot the finite-volume energies, 
scattering amplitude, and timelike pion form factor for all ensembles in 
Tab.~\ref{t:ens}.
The scattering amplitude and form factor tabulated here employ the 
truncation to $\ell\le 1$. 
The C101, D101, N401, N200, D200, J303 ensembles are tabulated in Tabs.~\ref{t:c101}, \ref{t:d101}, \ref{t:n401}, \ref{t:n200}, \ref{t:d200}, \ref{t:j303} and 
plotted in Figs.~\ref{f:c101}, \ref{f:d101},
\ref{f:n401}, \ref{f:n200}, \ref{f:d200}, \ref{f:j303}, respectively. 

\begin{table}
\centering 
\input{table_res_c101.tex}
	\caption{\label{t:c101}Results from the C101 ensemble.} 
\end{table}

\begin{table}
	\vspace{-2em}
\centering 
\input{table_res_d101.tex}
	\caption{\label{t:d101}Results from the D101 ensemble.}	
\end{table}

\begin{table}
\centering 
\input{table_res_n401.tex}
	\caption{\label{t:n401} Results from the N401 ensemble. The form factor is omitted for a single level in the $\Lambda(\boldsymbol{d}^2) = A_{1}^{+}(4)$ irrep for which a plateau could not be identified.} 
\end{table}

\begin{table}
\centering 
\input{table_res_n200.tex}
	\caption{\label{t:n200} Results from the N200 ensemble.} 
\end{table}

\begin{figure}[ht]
  \centering
	\includegraphics{tikz-figures/n200_ell1plot.tikz}
	\includegraphics{tikz-figures/n200_ffPlot_disp_nSubtr3.tikz}
	\includegraphics{tikz-figures/n200_dispFit_nSubtr3.tikz}
	\caption{\label{f:n200} Same as Fig.~\ref{f:n401} for the N200 ensemble.} 
\end{figure}

\begin{table}
\centering 
\input{table_res_d200.tex}
	\caption{\label{t:d200} Results from the D200 ensemble.} 
\end{table}

\begin{figure}[ht]
  \centering
	\includegraphics{tikz-figures/d200_ell1plot.tikz}
	\includegraphics{tikz-figures/d200_ffPlot_disp_nSubtr3.tikz}
	\hspace{-1.0em}\includegraphics{tikz-figures/d200_dispFit_nSubtr3.tikz}
  \caption{\label{f:d200} Same as Fig.~\ref{f:n401} for the D200 ensemble.}
\end{figure}

\begin{table}
\centering 
\input{table_res_j303.tex}
	\caption{\label{t:j303} Results from the J303 ensemble.} 
\end{table}

%% file: table_res_c101.tex
\begin{tabular}{cccccccc}
\toprule
$\boldsymbol{d}^{2}$ &  irrep.  &   level    &  $E_{\cm}/m_{\pi}$  & $p_{\rm cm}/m_{\pi}$ & $(p_{\rm cm}/m_{\pi})^3 \cot \delta_{1}$ & $\delta_{1}$ \\
\midrule
$0$   &    $T_{1u}^{+}$   & 0  & 3.148(23) & 1.477(37) & 2.99(40) &   31.0(2.5) \\ 
       &            & 1  & 3.564(33) & 2.176(58) & -3.34(42) &   136.1(4.6) \\ 
\midrule
$1$   &    $A_{1}^{+}$   & 0  & 2.2864(61) & 0.3069(70) & 7.5(1.7) &   1.29(24) \\ 
       &            & 1  & 3.318(32) & 1.752(53) & 1.18(31) &   63.0(5.0) \\ 
       &    $E^{+}$   & 0  & 3.211(34) & 1.578(54) & 2.13(31) &   43.0(2.7) \\ 
\midrule
$2$   &    $A_{1}^{+}$   & 0  & 2.4630(75) & 0.5165(93) & 6.4(1.0) &   3.32(45) \\ 
       &            & 1  & 3.426(37) & 1.934(64) & 0.40(24) &   81.6(4.7) \\ 
       &            & 2  & 3.821(26) & 2.650(50) & -5.6(1.3) &   142.4(6.7) \\ 
       &    $B_{1}^{+}$   & 0  & 3.310(34) & 1.739(57) & 2.34(30) &   44.4(2.5) \\ 
       &            & 1  & 3.783(22) & 2.579(42) & -8.3(6.7) &   153(28) \\ 
       &    $B_{2}^{+}$   & 0  & 2.692(12) & 0.812(16) & 5.59(80) &   7.45(85) \\ 
       &            & 1  & 3.409(40) & 1.905(68) & -0.39(20) &   98.4(4.6) \\ 
\midrule
$3$   &    $A_{1}^{+}$   & 0  & 2.5980(100) & 0.687(13) & 6.08(99) &   5.36(73) \\ 
       &            & 1  & 2.999(14) & 1.249(21) & 2.5(1.3) &   29(11) \\ 
       &            & 2  & 3.476(46) & 2.021(81) & -1.12(16) &   111.4(3.9) \\ 
       &    $E^{+}$   & 0  & 2.922(18) & 1.135(26) & 5.51(76) &   12.4(1.3) \\ 
       &            & 1  & 3.570(56) & 2.186(100) & -1.81(25) &   119.2(4.8) \\ 
\midrule
$4$   &    $A_{1}^{+}$   & 0  & 2.710(15) & 0.836(21) & 5.9(1.5) &   7.4(1.5) \\ 
       &            & 1  & 3.422(71) & 1.93(12) & -0.44(43) &   99.3(9.7) \\ 
       &    $E^{+}$   & 0  & 3.189(23) & 1.542(37) & 3.44(44) &   29.1(2.5) \\ 
       &            & 1  & 3.593(48) & 2.228(87) & -0.98(89) &   106(15) \\ 
\bottomrule
\end{tabular}

%% file: table_res_d101.tex
\begin{tabular}{cccccccc}
\toprule
$\boldsymbol{d}^{2}$ &  irrep.  &   level    &  $E_{\cm}/m_{\pi}$  & $p_{\rm cm}/m_{\pi}$ & $(p_{\rm cm}/m_{\pi})^3 \cot \delta_{1}$ & $\delta_{1}$ \\
\midrule
$0$   &    $T_{1u}^{+}$   & 0  & 2.7894(88) & 0.945(12) & 4.23(43) &   12.3(1.0) \\ 
       &            & 1  & 3.236(22) & 1.617(35) & 1.16(27) &   60.6(5.0) \\ 
       &            & 2  & 3.646(30) & 2.324(55) & -4.41(80) &   141.2(5.8) \\ 
\midrule
$1$   &    $A_{1}^{+}$   & 0  & 2.1880(38) & 0.1968(42) & 4.6(1.6) &   1.09(30) \\ 
       &            & 1  & 2.9555(99) & 1.184(15) & 4.42(47) &   16.2(1.5) \\ 
       &            & 2  & 3.324(20) & 1.761(34) & 1.13(24) &   64.1(4.1) \\ 
       &            & 3  & 3.564(24) & 2.175(42) & -6.3(7.3) &   153(31) \\ 
       &            & 4  & 3.674(20) & 2.374(36) & -4.04(83) &   137.8(6.3) \\ 
       &    $E^{+}$   & 0  & 2.912(12) & 1.121(17) & 4.32(47) &   15.4(1.3) \\ 
       &            & 1  & 3.358(18) & 1.820(31) & 0.22(18) &   84.9(4.2) \\ 
       &            & 2  & 3.740(26) & 2.496(49) & -5.76(94) &   145.6(5.0) \\ 
\midrule
$2$   &    $A_{1}^{+}$   & 0  & 2.3229(49) & 0.3490(57) & 4.7(1.2) &   2.49(55) \\ 
       &            & 1  & 3.068(11) & 1.353(17) & 2.89(32) &   28.5(2.4) \\ 
       &            & 2  & 3.297(19) & 1.717(32) & 0.15(19) &   86.2(4.8) \\ 
       &            & 3  & 3.693(16) & 2.409(29) & -2.6(1.2) &   125(13) \\ 
       &            & 4  & 3.750(22) & 2.515(41) & -9.0(4.2) &   156.0(8.7) \\ 
       &    $B_{1}^{+}$   & 0  & 2.988(13) & 1.233(19) & 3.55(30) &   21.1(1.2) \\ 
       &            & 1  & 3.149(11) & 1.479(18) & 1.2(3.6) &   56(60) \\ 
       &            & 2  & 3.535(30) & 2.124(54) & -2.19(21) &   125.2(3.5) \\ 
       &    $B_{2}^{+}$   & 0  & 2.4392(51) & 0.4875(62) & 6.82(97) &   2.86(37) \\ 
       &            & 1  & 3.114(11) & 1.424(18) & 2.73(24) &   31.9(2.0) \\ 
       &            & 2  & 3.346(20) & 1.798(33) & -0.03(20) &   90.8(4.8) \\ 
       &            & 3  & 3.620(15) & 2.276(27) & -13(12) &   165(14) \\ 
       &            & 4  & 3.779(24) & 2.569(45) & -12.2(4.3) &   161.4(6.1) \\ 
\midrule
$3$   &    $A_{1}^{+}$   & 0  & 2.4411(78) & 0.4897(95) & 9.3(5.9) &   2.1(1.1) \\ 
       &            & 1  & 2.6267(67) & 0.7250(88) & 4.2(2.5) &   8.4(4.2) \\ 
       &            & 2  & 3.281(24) & 1.692(40) & 1.00(16) &   65.6(2.7) \\ 
       &            & 3  & 3.742(19) & 2.501(35) & -9.7(4.5) &   157.9(9.6) \\ 
       &            & 4  & 3.878(22) & 2.760(44) & -9.0(5.8) &   152.9(5.6) \\ 
       &            & 5  & 3.938(28) & 2.876(55) & -9.4(1.5) &   152.6(4.3) \\ 
       &    $E^{+}$   & 0  & 2.6031(68) & 0.6940(89) & 6.04(69) &   5.47(55) \\ 
       &            & 1  & 3.369(21) & 1.838(36) & 1.16(15) &   65.1(2.3) \\ 
       &            & 2  & 3.697(16) & 2.417(30) & -9(12) &   157(42) \\ 
       &            & 3  & 3.862(16) & 2.728(31) & 2(10) &   69(65) \\ 
\midrule
$4$   &    $A_{1}^{+}$   & 0  & 2.494(15) & 0.555(18) & 2.14(70) &   11.0(2.9) \\ 
       &            & 1  & 3.171(13) & 1.514(21) & 2.34(18) &   38.5(1.7) \\ 
       &            & 2  & 3.566(20) & 2.178(35) & -1.62(23) &   116.7(3.6) \\ 
       &            & 3  & 3.959(19) & 2.919(38) & -8.0(1.9) &   148.2(6.5) \\ 
       &    $E^{+}$   & 0  & 2.7979(94) & 0.957(13) & 4.39(68) &   12.0(1.6) \\ 
       &            & 1  & 3.222(13) & 1.595(21) & 3.27(30) &   31.6(2.0) \\ 
       &            & 2  & 3.406(16) & 1.900(28) & -0.6(1.1) &   102(22) \\ 
       &            & 3  & 3.638(19) & 2.309(34) & -2.47(28) &   125.2(3.4) \\ 
       &            & 4  & 3.980(21) & 2.959(43) & -9.9(2.0) &   152.9(5.0) \\ 
\bottomrule
\end{tabular}

%% file: table_res_n401.tex
\begin{tabular}{ccccccccc}
\toprule
$\boldsymbol{d}^{2}$ &  irrep.  &   level    &  $E_{\cm}/m_{\pi}$  & $p_{\rm cm}/m_{\pi}$ & $(p_{\rm cm}/m_{\pi})^3 \cot \delta_{1}$ & $\delta_{1}$ & $|F_{\pi}|$ \\
\midrule
$0$   &    $T_{1u}^{+}$   & 0  & 2.714(22) & 0.841(30) & 0.281(64) &   70.0(3.3) &  17.05(69)   \\ 
       &            & 1  & 3.205(13) & 1.568(21) & -5.18(57) &   159.2(2.4) &  3.518(53)   \\ 
\midrule
$1$   &    $A_{1}^{+}$   & 0  & 2.2248(24) & 0.2375(27) & 2.89(19) &   2.29(12) &  3.019(29)   \\ 
       &            & 1  & 2.798(22) & 0.957(31) & -0.481(39) &   117.2(2.9) &  12.42(77)   \\ 
       &    $E^{+}$   & 0  & 2.775(20) & 0.925(28) & 0.269(48) &   73.2(2.2) &  13.92(99)   \\ 
\midrule
$2$   &    $A_{1}^{+}$   & 0  & 2.3597(55) & 0.3920(65) & 1.98(18) &   7.05(47) &  4.257(53)   \\ 
       &            & 1  & 2.836(23) & 1.010(33) & -0.958(48) &   133.3(2.9) &  10.38(60)   \\ 
       &    $B_{1}^{+}$   & 0  & 2.748(21) & 0.888(28) & 0.094(38) &   83.6(2.3) &  15.8(1.1)   \\ 
       &    $B_{2}^{+}$   & 0  & 2.509(11) & 0.574(14) & 1.55(22) &   15.7(1.6) &  6.49(15)   \\ 
       &            & 1  & 2.907(22) & 1.113(33) & -1.316(79) &   138.2(2.8) &  8.26(30)   \\ 
\midrule
$3$   &    $A_{1}^{+}$   & 0  & 2.4699(91) & 0.525(11) & 1.87(23) &   11.5(1.0) &  6.07(12)   \\
       &            & 1  & 2.806(15) & 0.969(20) & -0.01(72) &   90(36) &  19.8(3.3)   \\
       &            & 2  & 2.950(34) & 1.176(51) & -2.03(29) &   147.8(5.2) &  6.32(60)   \\
       &    $E^{+}$   & 0  & 2.627(31) & 0.725(41) & 0.96(26) &   32.7(4.6) &  11.26(98)   \\
       &            & 1  & 3.041(33) & 1.313(50) & -2.99(38) &   153.3(4.1) &  5.78(21)   \\
\midrule
$4$   &    $A_{1}^{+}$   & 0  & 2.582(26) & 0.666(34) & 2.25(91) &   13.6(3.8) &  12.3(1.0)   \\ 
       &            & 1  & 2.834(23) & 1.008(32) & -2.40(27) &   157.1(3.2) &  --- \\ 
       &    $E^{+}$   & 0  & 2.722(39) & 0.853(53) & 0.23(11) &   73.5(6.0) &  16.5(1.4)   \\ 
       &            & 1  & 3.194(16) & 1.550(26) & -4.24(81) &   155.5(4.5) &  4.360(92)   \\ 
\bottomrule
\end{tabular}

%% file: table_res_n200.tex
\begin{tabular}{ccccccccc}
\toprule
$\boldsymbol{d}^{2}$ &  irrep.  &   level    &  $E_{\cm}/m_{\pi}$  & $p_{\rm cm}/m_{\pi}$ & $(p_{\rm cm}/m_{\pi})^3 \cot \delta_{1}$ & $\delta_{1}$ & $|F_{\pi}|$ \\
\midrule
$0$   &    $T_{1u}^{+}$   & 0  & 2.749(27) & 0.889(37) & -0.077(41) &   95.3(3.1) &  15.3(1.1)   \\ 
\midrule
$1$   &    $A_{1}^{+}$   & 0  & 2.2654(24) & 0.2830(27) & 2.78(10) &   3.104(68) &  3.308(36)   \\ 
       &            & 1  & 2.874(22) & 1.065(32) & -1.176(36) &   136.9(2.1) &  8.88(42)   \\ 
       &    $E^{+}$   & 0  & 2.756(35) & 0.898(48) & -0.179(40) &   101.8(3.5) &  16.1(1.5)   \\ 
\midrule
$2$   &    $A_{1}^{+}$   & 0  & 2.4215(66) & 0.4659(80) & 2.22(16) &   8.14(36) &  4.94(12)   \\ 
       &            & 1  & 2.913(29) & 1.121(43) & -2.059(96) &   150.0(2.6) &  7.04(48)   \\ 
       &    $B_{1}^{+}$   & 0  & 2.676(51) & 0.791(68) & -0.311(32) &   113.9(4.8) &  16.7(1.5)   \\ 
       &    $B_{2}^{+}$   & 0  & 2.613(10) & 0.707(14) & 1.20(10) &   26.4(1.3) &  10.87(47)   \\ 
       &            & 1  & 3.089(24) & 1.385(37) & -2.99(22) &   151.4(2.6) &  4.67(16)   \\ 
\midrule
$3$   &    $A_{1}^{+}$   & 0  & 2.482(18) & 0.540(22) & 1.26(17) &   17.5(1.2) &  7.29(21)   \\ 
       &            & 1  & 2.959(39) & 1.189(58) & -2.51(52) &   152.7(6.8) &  5.85(45)   \\ 
       &            & 2  & 3.176(20) & 1.522(32) & -5.2(1.4) &   160.3(5.0) &  3.60(51)   \\ 
       &    $E^{+}$   & 0  & 2.56(11) & 0.63(13) & 0.14(21) &   74(16) &  12.4(1.7)   \\ 
\midrule
$4$   &    $A_{1}^{+}$   & 0  & 2.582(69) & 0.667(89) & 1.22(66) &   24.0(6.4) &  10.0(1.9)   \\ 
       &    $E^{+}$   & 0  & 2.778(96) & 0.93(13) & -0.10(15) &   96(11) &  14.8(3.6)   \\ 
\bottomrule
\end{tabular}

%% file: table_res_d200.tex
\begin{tabular}{ccccccccc}
\toprule
$\boldsymbol{d}^{2}$ &  irrep.  &   level    &  $E_{\cm}/m_{\pi}$  & $p_{\rm cm}/m_{\pi}$ & $(p_{\rm cm}/m_{\pi})^3 \cot \delta_{1}$ & $\delta_{1}$ & $|F_{\pi}|$ \\
\midrule
$0$   &    $T_{1u}^{+}$   & 0  & 3.444(22) & 1.965(38) & 6.39(86) &   23.3(2.3) &  4.655(98)   \\ 
       &            & 1  & 3.980(75) & 2.96(15) & -2.5(1.3) &   116(13) &  7.55(37)   \\ 
\midrule
$1$   &    $A_{1}^{+}$   & 0  & 2.3385(45) & 0.3672(53) & 10.1(1.3) &   1.26(14) &  1.656(16)   \\ 
       &            & 1  & 3.633(30) & 2.300(55) & 2.53(60) &   54.1(5.4) &  6.81(22)   \\ 
       &    $E^{+}$   & 0  & 3.548(32) & 2.147(57) & 4.64(68) &   34.1(2.8) &  5.81(18)   \\ 
\midrule
$2$   &    $A_{1}^{+}$   & 0  & 2.5386(65) & 0.6111(83) & 9.0(1.2) &   3.02(33) &  1.908(18)   \\ 
       &            & 1  & 3.748(40) & 2.512(75) & 1.05(47) &   75.3(5.6) &  7.77(27)   \\ 
       &    $B_{1}^{+}$   & 0  & 3.637(26) & 2.308(48) & 4.32(41) &   39.1(1.9) &  6.81(21)   \\ 
       &    $B_{2}^{+}$   & 0  & 2.851(11) & 1.033(16) & 8.5(1.3) &   7.06(94) &  2.369(31)   \\ 
       &            & 1  & 3.794(43) & 2.598(81) & 0.36(43) &   85.1(5.6) &  8.50(19)   \\ 
\midrule
$3$   &    $A_{1}^{+}$   & 0  & 2.701(11) & 0.824(14) & 11.2(2.5) &   3.83(74) &  2.056(21)   \\ 
       &            & 1  & 3.213(15) & 1.580(24) & 4.6(4.0) &   23(16) &  4.12(67)   \\ 
       &            & 2  & 3.886(91) & 2.78(18) & -0.83(52) &   100.2(7.1) &  8.79(38)   \\ 
       &    $E^{+}$   & 0  & 3.128(19) & 1.446(29) & 8.7(1.6) &   11.4(1.7) &  3.137(40)   \\ 
       &            & 1  & 3.983(84) & 2.97(17) & -1.79(62) &   109.3(7.6) &  8.13(64)   \\ 
\midrule
$4$   &    $A_{1}^{+}$   & 0  & 2.839(19) & 1.015(27) & 14.1(6.5) &   4.2(1.5) &  2.354(27)   \\ 
       &    $E^{+}$   & 0  & 3.499(37) & 2.061(64) & 8.4(3.2) &   19.3(5.4) &  4.94(16)   \\ 
\bottomrule
\end{tabular}

%% file: table_res_j303.tex
\begin{tabular}{ccccccccc}
\toprule
$\boldsymbol{d}^{2}$ &  irrep.  &   level    &  $E_{\cm}/m_{\pi}$  & $p_{\rm cm}/m_{\pi}$ & $(p_{\rm cm}/m_{\pi})^3 \cot \delta_{1}$ & $\delta_{1}$ & $|F_{\pi}|$ \\
\midrule
$0$   &    $T_{1u}^{+}$   & 0  & 3.044(28) & 1.317(42) & 0.50(11) &   71.6(3.0) &  11.36(30)   \\ 
\midrule
$1$   &    $A_{1}^{+}$   & 0  & 2.3048(48) & 0.3280(55) & 4.32(36) &   2.49(15) &  2.460(25)   \\ 
       &            & 1  & 3.211(29) & 1.578(47) & -0.802(87) &   112.0(3.0) &  9.12(41)   \\ 
       &    $E^{+}$   & 0  & 3.082(43) & 1.374(66) & 0.32(14) &   78.6(4.0) &  11.51(46)   \\ 
\midrule
$2$   &    $A_{1}^{+}$   & 0  & 2.4890(79) & 0.5488(98) & 3.95(40) &   5.87(43) &  3.291(40)   \\ 
       &            & 1  & 3.236(48) & 1.617(77) & -1.70(13) &   129.6(4.2) &  8.10(67)   \\ 
       &    $B_{1}^{+}$   & 0  & 3.001(64) & 1.251(96) & -0.05(13) &   91.9(5.4) &  11.63(64)   \\ 
       &    $B_{2}^{+}$   & 0  & 2.755(27) & 0.898(38) & 2.66(65) &   17.7(3.0) &  5.55(12)   \\ 
       &            & 1  & 3.199(83) & 1.56(13) & -3.48(83) &   150.8(8.5) &  7.9(1.2)   \\ 
\midrule
$3$   &    $A_{1}^{+}$   & 0  & 2.599(19) & 0.688(25) & 2.90(51) &   11.1(1.3) &  4.197(67)   \\ 
       &            & 1  & 3.165(36) & 1.505(57) & -0.7(1.6) &   109(40) &  10.0(2.8)   \\ 
       &            & 2  & 3.30(11) & 1.71(18) & -5.6(3.1) &   158(30) &  7.5(4.1)   \\ 
       &    $E^{+}$   & 0  & 2.936(28) & 1.155(42) & 1.92(34) &   32.9(3.3) &  9.12(41)   \\ 
       &            & 1  & 3.550(61) & 2.15(11) & -5.22(85) &   148.9(5.8) &  3.89(18)   \\ 
\midrule
$4$   &    $A_{1}^{+}$   & 0  & 2.694(30) & 0.814(41) & 2.48(59) &   16.5(2.5) &  5.02(12)   \\ 
       &            & 1  & 3.281(73) & 1.69(12) & -2.95(46) &   143.3(7.1) &  7.03(78)   \\ 
       &    $E^{+}$   & 0  & 3.049(64) & 1.324(98) & 0.37(24) &   76.3(6.9) &  11.94(54)   \\ 
       &            & 1  & 3.645(46) & 2.321(83) & -27(38) &   173(84) &  2.598(91)   \\ 
\bottomrule
\end{tabular}